%% file: main.tex
\documentclass[apj]{emulateapj}
\usepackage{amsfonts}
\usepackage{amsmath}
\usepackage{graphicx}
\usepackage{natbib}
\usepackage{apjfonts}
\usepackage{subfigure}
\usepackage{color}
\usepackage{longtable}

\usepackage{perpage} 
\MakePerPage{footnote} 

\usepackage[caption=false]{subfig}

\newcommand{\angstrom}{\mbox{\normalfont\AA}}

\newcommand{\csigma}{$11\sigma$~}
\newcommand{\gsigma}{$4.5\sigma$~}

\shorttitle{Brout and Scolnic 2020}
\shortauthors{Brout and Scolnic}

\begin{document}

\title{It's Dust: Solving the Mysteries of the Intrinsic Scatter and Host-Galaxy Dependence of Standardized Type Ia Supernova Brightnesses}

\def\andname{}

\author{Dillon  Brout \altaffilmark{1,2} \& Daniel Scolnic \altaffilmark{3} \vspace{.05in}}

\affil{$^{1}$ Department of Physics and Astronomy, University of Pennsylvania, Philadelphia, PA 19104, USA}
\affil{$^{2}$ NASA Einstein Fellow}
\affil{$^{3}$ Department of Physics, Duke University, Durham, NC 27708, USA \vspace{.1in}}
\email{dbrout@physics.upenn.edu}
\email{daniel.scolnic@duke.edu}




\accepted{to The Astrophysical Journal}

\begin{abstract}

The use of Type Ia Supernovae (SNe~Ia) as cosmological tools has motivated significant effort to: understand what drives the intrinsic scatter of SN~Ia distance modulus residuals after standardization, characterize the distribution of SN~Ia colors, and explain why properties of the host galaxies of the SNe correlate with SN~Ia distance modulus residuals. We use a compiled sample of $\sim1450$ spectroscopically confirmed, photometric light-curves of SN~Ia and propose a solution to these three problems simultaneously that also explains an empirical \csigma detection of the dependence of Hubble residual scatter on SN~Ia color. We introduce a physical model of color where intrinsic SN~Ia colors with a relatively weak correlation with luminosity are combined with extrinsic dust-like colors ($E(B-V)$) with a wide range of extinction parameter values ($R_V$). This model captures the observed trends of Hubble residual scatter and indicates that the dominant component of SN~Ia intrinsic scatter is from variation in $R_V$. We also find that the recovered $E(B-V)$ and $R_V$ distributions differ based on global host-galaxy stellar mass and this explains the observed correlation ($\gamma$) between mass and Hubble residuals seen in past analyses as well as an observed \gsigma dependence of $\gamma$ on SN~Ia color. This finding removes any need to prescribe different intrinsic luminosities to different progenitor systems. Finally we measure biases in the equation-of-state of dark energy ($w$) up to $|\Delta w|=0.04$ by replacing previous models of SN color with our dust-based model; this bias is larger than any systematic uncertainty in previous SN~Ia cosmological analyses.

\end{abstract}
\keywords{supernovae, cosmology}
\section{Introduction}
\label{Sec:intro}

 Studies in the last decade of research in cosmology with Type Ia supernovae (SNe~Ia) have forewarned that the measurements of the equation-of-state of dark energy $w$ will soon hit a systematic floor. Yet, such measurements (B14:~\citealt{Betoule2014}, S18:~\citealt{Scolnic18}, B19b:~\citealt{Brout18-SYS}, \citealt{Jones19}) continually reach better levels of both statistical and systematic precision. This is due to the improvement of systematic uncertainties in survey and camera design, but also due to the possibility afforded from significantly larger samples to understand systematics in the analysis. In the most recent analyses (S18, B19b), it has been found that systematic uncertainties in understanding the intrinsic scatter of standardized SN~Ia brightnesses is of a similar level or larger than uncertainties due to external, photometric calibration. As calibration uncertainties have been dominant in past systematic error budgets, this moment marks a transition from a need to understand external issues independent of the supernovae to a need to also better understand SN~Ia physics.
 
 With current cosmological analyses of SNe~Ia requiring mmag-level control of systematics, uncertainty over how to understand the intrinsic scatter of standardized SN~Ia brightnesses, which is on the 0.1 mag level, is problematic. Practically, intrinsic scatter is measured as the excess scatter of SN~Ia distance residuals to a best-fit cosmology after accounting for measurement noise.
 A holistic understanding of SN~Ia intrinsic scatter and its underlying characterization has remained elusive, but its size has been found to depend on a wide variety of measurement components: redshift (e.g., B14), wavelength range of the photometric observations (e.g., \citealp{mandel11}), host-galaxy properties (e.g., \citealt{Uddin17}), and spectroscopic features (e.g., \citealp{Fakhouri15}). 
 Furthermore, \cite{Scolnic16} showed that the relative amounts of chromatic versus achromatic components of the intrinsic scatter models were directly linked to the intrinsic SN~Ia color population and reddening law; however, this study was unable to discriminate between different models.

After the discovery of the accelerating universe \citep{Riess98,Perlmutter99}, there were two commonly used light-curve fitters: MLCS2k2 \citep{Jha2007} and SALT2 \citep{Guy2010}, that diverged in their approach to color and intrinsic scatter. MLCS2k2 attempted to model color based on dust with the possibility that each SN could have its own extinction law, and assumed that a large amount of the intrinsic scatter was in color. The SALT2 model, on the other hand, was agnostic to any physical properties of the SN color and its relation to the intrinsic scatter. Cosmological analyses have since favored the SALT2 model due to its native spectral-model to account for k-corrections and updated calibration, and it has been used in most recent cosmology analyses including the Joint Light-Curve Analysis (JLA: B14), Pantheon (S18), the Dark Energy Survey 3 Year Sample (DES3YR: \citealt{Brout18-SMP}, B19a), and the Foundation + Pan-STARRS1 photometric analysis (\citealt{Jones19}). However, despite the fact that MLCS2k2 has not been used in recent cosmological analyses, papers such as \cite{ScolnicMW,Scolnic18,Mandel17} have attempted to bridge the gap between SALT2 and MLCS2k2 methods by modeling a connection between the underlying population of color, dust, and reddening laws.

Still, SN~Ia analyses that attempt to model dust using a cosmological sample have typically made the simplistic assumption that there is a single total-to-selective extinction parameter, $R_V$, that can be fixed at a single number. $R_V$ is defined as $A_V/(A_B-A_V)$, where $A_V$ is the extinction in the V ($\lambda V \sim 5500~\angstrom$) band, and $A_B$ is the extinction in the blue ($\lambda V \sim 4400~\angstrom$) band. As $R_V$ varies for different dust grain sizes and composition, and galaxies have different dust properties, it is well known that different galaxies and different regions within galaxies exhibit a wide range of $R_V$ values. In fact, while the Milky Way galaxy has an $R_V$ on average $\sim3.1$, it has a distribution of at least $\sigma_{R_V}=0.2$ \citep{Schlafly16}. Additionally, different parts of the LMC and SMC have been found to have $R_V$ values with a range of $R_V\sim2-5$ \citep{Gao2013,Yanchulova17}. Furthermore, \cite{Salim18} study the dust attenuation curves of 230,000 individual galaxies in the local universe, using GALEX, SDSS, and WISE photometry calibrated on the Herschel ATLAS, and they find quiescent galaxies, which are typically high-mass, have a mean $R_V=2.61$ and star-forming galaxies, which are lower-mass on average, have a mean $R_V=3.15$.

$R_V$ has also been measured through large SN sample statistics and detailed studies of individual SNe, though often with varying sets of assumptions. \cite{Cikota16} compiled 13 various studies of SN~Ia samples from the literature which determined a range of $R_V$ values from $\sim1$ to $\sim 3.5$. \cite{Cikota16} itself determined $R_V$ from nearby SNe and for 21 SNe Ia observed in Sab-Sbp galaxies and 34 SNe in Sbc-Scp they find $R_V=2.71\pm1.58$ and $R_V = 1.70 \pm 0.38$ respectively. While so many past analyses have recovered $R_V<2$ for studies of individual SNe (e.g. \citealp{XWang05,Krisciunas06}), these were often SNe~Ia with high $E(B-V)$, and it was postulated $R_V$ may decrease with $E(B-V)$. However, \cite{Nobili_2008} found from a sample of modestly reddened ($E(B-V)< 0.25$ mag) SNe~Ia, a small value of $R_V\sim1$ and more recently, \cite{Amanullah15} analyzed high-quality UV-NIR spectra of 6 SNe and found that SNe with high reddening indicated $R_V$'s ranging from $\sim1.4$ to $\sim2.8$ and SNe with low amounts of reddening also indicated $R_V$'s of $\sim1.4$ and $\sim2.8$. Importantly, \cite{Amanullah15} stressed that the observed diversity in $R_V$ is not accounted for in analyses that measure the cosmological expansion of the universe.

Since the low $R_V$ values ($<2$) are not found in studies of the Milky Way, this has motivated various SN~Ia studies to ascribe the dust to circumstellar dust around the progenitor at the time of the explosion \citep{Wang05,Goobar08}. However, an alternative interpretation could be that the low $R_V$ values are caused by dust in the interstellar medium \citep{Phillips13}. This understanding has been supported by \citet{Bulla18a,Bulla18b}, which constrained the location of the dust that caused the reddening in the SN~Ia spectra to be, for the majority of the SNe that they observed, on scales of the interstellar medium, rather than circumstellar surroundings. This could be due to cloud-cloud collisions induced by the SN radiation pressure \citep{Hoang17} which produce small dust grains \citep{Gao15, Nozawa16}. 

While accounting for dust remains a challenge for current and future photometric cosmology analyses, this pursuit has often been done in parallel to the search for correlations between measured supernova luminosity after standardization and host-galaxy properties. Global and local properties of SN~Ia host galaxies such as stellar mass, star formation rate (SFR), stellar population age, and metallicity have all been shown to correlate with the distance modulus residuals after standardization (\citealt{Hicken09a,Sullivan2010,Lampeitl2010,Childress2013,Rose19}). This correlation is often parameterized as a step function in host-galaxy stellar mass and is now commonplace in SN~Ia cosmology analyses despite the lack of understanding of its physical underpinning or convincing evidence for exactly which host-galaxy property is most influential on SN~Ia luminosity (e.g. \citealt{Jones18global,scolnicsiblings}). To explain this correlation, recent studies have suggested a potential relation between the luminosity of the SN and the progenitor, which can be related to the age of the galaxy, or the local environment of the galaxy \citep{Childress2013,Rigault2013,Roman18}. However, as the aforementioned galaxy properties are all directly linked to dust properties, it is likely that the lack of dust modeling in SN~Ia cosmology is related to the correlations between host galaxy properties and standardized luminosities.

In this analysis, we show that there are clear limitations in SN~Ia standardization techniques with a single color luminosity correlation, but that these limitations can be addressed by inclusion of dust modeling with variation in $R_V$. This paper relies heavily  on the work of \cite{Mandel17}, which follows closer to the framework of MLCS2k2 and developed a hierarchical Bayesian model to build a more rich understanding of SN color. \cite{Mandel17} only used low-redshift data, did not account for selection effects, and assumed a fixed $R_V$ extinction parameter; here 
we use a much larger dataset across a wide redshift range and use survey simulations to forward-model what is done in \cite{mandel11}, though with additional features to explain discrepancies seen between simulations and data.

In Section 2, we present the data compilation, light-curve fitting and discrepancies between the data and a simple understanding of SN color. In Section 3, we discuss how to differentiate between past models of SN color and our new dust-based color model. In Section 4, we show how the new model can explain the commonly seen correlation between distance modulus residuals and host-galaxy properties. In Section 5, we assess the impact on recovered cosmological parameters, and in Sections 6 \& 7, we discuss further studies and conclusions.

\section{Data Sample, Distance Moduli, and Description of SN~Ia Colors}

\begin{figure}
\centering
\includegraphics[width=0.49\textwidth]{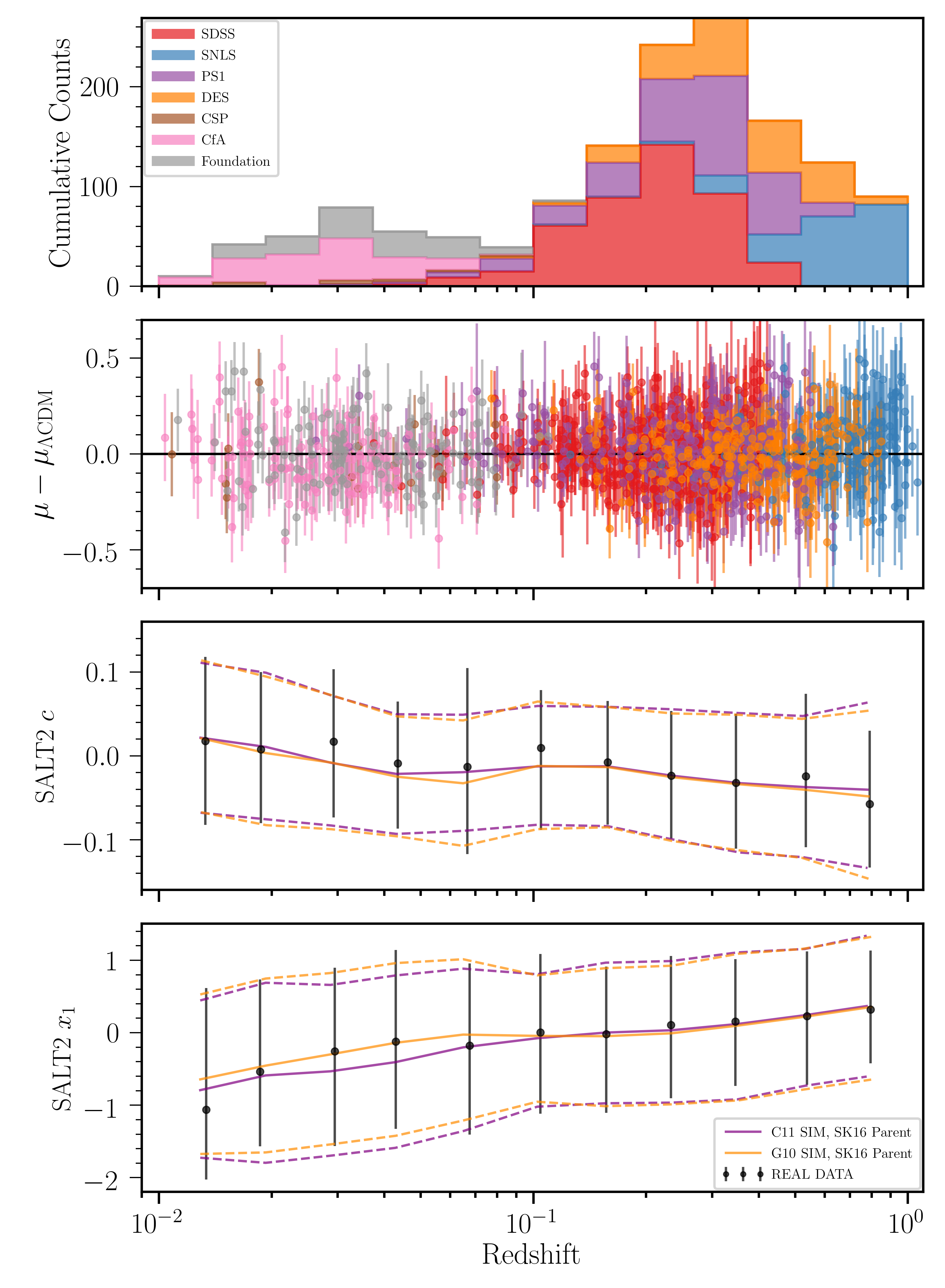}
\caption{Top: Stacked redshift histograms of each of the samples analyzed. Second: Hubble Diagram residuals relative to flat $\Lambda$CDM cosmology with $w=-1$. Third: Mean and 68\% intervals for the measured SALT2 color from the data, shown in blue points. Predictions from survey simulations shown in purple for simulations with the C11+SK16 model and orange for simulations with the G10+SK16 scatter model. Bottom: Same as third panel, but for the measured SALT2 stretch $x_1$.}

\label{figone}
\end{figure}

\subsection{Data}
We use a compilation of publicly available, spectroscopically classified, photometric light curves of SNe Ia that have been used in past cosmological analyses and that have been calibrated to the SuperCal system \citep{Scolnic2015}. The low-redshift (low-z) SNe used here are made up of, in part, by those used in B19b which are from CSP \citep{Stritzinger2010} and CfA3-4 \citep{Hicken09b,Hicken09a,Hicken12}. At low-z, we also include the recently released 180 low-z SNe from the Foundation sample \citep{Foley17}. At high-z, we include SNe from PS1 \citep{Rest14,Scolnic18}, SDSS \citep{Sako11} and SNLS (B14) as was done in the Pantheon analysis. Finally, we include data from the recently released DES 3-year sample \citep{Brout18-SMP}, hereafter DES3YR. The redshift distribution of SNe~Ia used in this work can be found in the top panel of Figure \ref{figone}.

This analysis relies largely on the host galaxy mass estimates provided by past analyses. We adopt the same masses released in the Pantheon sample, and references therein, for SDSS, PS1, SNLS, CSPDR2, and CfA. For DES3YR masses, we use the updated masses provided by \cite{Smith2020,Wiseman2020}. For the Foundation sample, we utilize masses derived in \cite{Jones18}.

\subsection{Light-curve fits and Distance Modulus Determination}
We fit the SNe with the SALT2 model as presented in \cite{Guy2010} and updated in B14. In SALT2, the SN~Ia flux at phase ($p$) and wavelength ($\lambda$) is given as
\begin{align} 
\label{saltmodel}
\begin{split}
F(\rm{SN}, p, \lambda) = x_{0} &\times\left[M_{0}(p, \lambda)+x_{1} M_{1}(p, \lambda)+\ldots\right] \\
&\times \exp [c C L(\lambda)],
\end{split}
\end{align}
where the parameter $x_0$ describes the overall amplitude of the light-curve, $x_1$ describes the observed light-curve stretch, and $c$ describes the observed color of each SN. 
$M_0$, $M_1$, $CL$ are global model parameters of all SNe~Ia: $M_0$ represents the average spectral sequence (SED); $M_1$
is the SED variability; and $CL$ is the average color correction law. The light-curve fits assume \cite{Fitzpatrick99} for Milky Way reddening. The mean observed $c$ and $x_1$ for the data, binned over redshift, is shown in the bottom panels of Figure \ref{figone}.

Distances are inferred following the Tripp estimator \citep{Tripp98}. The distance modulus ($\mu$) to each candidate SN~Ia is obtained by:\
\begin{equation}
\label{eq:tripp}
    \mu = m_B + \alpha_{\rm SALT2} x_1 - \beta_{\rm SALT2} c - M 
\end{equation}
where $m_B$ is peak-brightness based off of the light-curve amplitude (log$_{10}(x_0)$) and where $M$ is the absolute magnitude of a SN~Ia with $x_1=c=0$. $\alpha_{\rm SALT2}$ and $\beta_{\rm SALT2}$ are the correlation coefficients that standardize the SNe~Ia and are determined following \cite{Marriner11}, in a similar process to what is done in B14. \cite{Marriner11} minimize a $\chi^2$ expression that depends on the Hubble residuals after applying the Tripp estimator (see Eq.~\ref{eq:tripp}) and normalize residuals by the quadrature sum of the measurement uncertainties and intrinsic scatter. The method separates the sample into redshift bins in order to remove the cosmological dependence of the fitting procedure. The procedure iterates to determine the intrinsic scatter $\sigma_{int}$, $\alpha$, $\beta$ and the resultant distance modulus values.

In recent analyses with the Tripp estimator (S18, B19b), there is often additional additive terms $\delta_{\rm bias}$, the correction for distance biases calculated from survey simulations and $\delta_{\gamma}$, the correction due to the host-galaxy mass correlation; these additional corrections are not applied because new treatments for both of these terms are introduced in following sections. 

Distance uncertainties are computed from the uncertainties in the light-curve fit parameters and their covariance ($C$):
\begin{multline}
  \label{sigmastat}
   \sigma_\mu^2 = C_{m_B,m_B} + \alpha_{\rm SALT2}^2 C_{x_1,x_1} + \beta_{\rm SALT2}^2 C_{c,c} + 2\alpha_{\rm SALT2} C_{m_B,x_1} - \\ 2\beta_{\rm SALT2} C_{m_B,c} -  2\alpha_{\rm SALT2}\beta_{\rm SALT2} C_{x_1,c} + 
 \sigma^2_{\rm vpec} + \sigma^2_z + \sigma^2_{\rm lens} + \sigma_{\rm int}^2~,
\end{multline}
where $\sigma_{\rm vpec}$ is the distance modulus uncertainty due to peculiar velocities (250 km/s), $\sigma_{z}$ is the distance modulus uncertainty due to the measured redshift uncertainty, $\sigma_{\rm lens}$ is the additional uncertainty from weak gravitational lensing ($0.055z$), and $\sigma_{\rm int}$ is determined such that the reduced $\chi^2$ relative to a best fit cosmology is 1.

 Typical selection cuts are applied on the observed data sample as was done in B19b: we require fitted color uncertainty $< 0.05$, fitted stretch uncertainty $< 1$, fitted light-curve peak date uncertainty $< 2$, light-curve fit probability (from SNANA) $> 0.01$, and Chauvenaut's criterion is applied to distance modulus residuals, relative to the best fit cosmological model, at 3.5$\sigma$. In total, after selection cuts, there are 1445 SNe in this sample.

\subsection{Key Pillars of the Complexity of the Colors of SNe~Ia}
\label{sec:pillars}

The complexity of the SN~Ia color model is readily apparent after a simple Tripp standardization. Here, three critical features are presented in the observed dataset that must be explained by models of SN~Ia color and intrinsic scatter. 
\begin{itemize}
    \item The distribution of observed SN~Ia colors is shown in the top of Fig. \ref{fig:cartoonmetric}. There is a clear asymmetry, with an excess of red SNe in comparison with blue SNe, that is inconsistent with a symmetric Gaussian distribution. 

    \item The relation between the root-mean-square (RMS) scatter of distance modulus residuals (with mean residual removed in each bin) as a function of SN~Ia color is shown in the middle panel of Fig. \ref{fig:cartoonmetric}. There is a \csigma dependence relative to a flat line, where the redder SNe~Ia ($c>0.1$) exhibit nearly twice as much scatter ($\sim$0.18~mag) as the bluest SNe ($c<-0.1$), which exhibit $\sim$0.1~mag scatter. This effect remains if any single survey is removed from the sample. 
    
    \item The relation between Hubble residual binned distance biases and SN~Ia color is shown in the bottom panel of Fig. \ref{fig:cartoonmetric}. There is a $\sim7.8\sigma$ dependence relative to a straight line.  As shown in Fig \ref{fig:cartoonmetric}, the recovered $\beta_{\rm SALT2}$ of the data is $3.05\pm0.06$.

\end{itemize}
The relation of increased scatter as a function of color has not been analyzed in a previous analysis.  This paper is motivated by quantifying these observed features and building a model that can address all of them simultaneously.

\subsection{Using Survey Simulations to Evaluate SN~Ia Color and Intrinsic Scatter Models}
\label{sec:fittingmethod}

For every model presented in this paper, 100 realizations of dataset-sized simulations are run. \texttt{SNANA} \citep{Kessler2009} is used to simulate realistic samples of SNe~Ia. These simulations account for observing cadence, observing conditions, noise properties, selection effects, cosmological effects, and astrophysical effects. A general description of the simulation methodology can be found in \citet{Kessler18} and the survey specific simulation details for SDSS and SNLS are described in \cite{Kessler13}; PS1, CSP, and CfA are described in S18; DES3YR is described in B19b and Foundation is described in \cite{Jones18}. 

\begin{figure}
\centering
\includegraphics[width=0.48\textwidth]{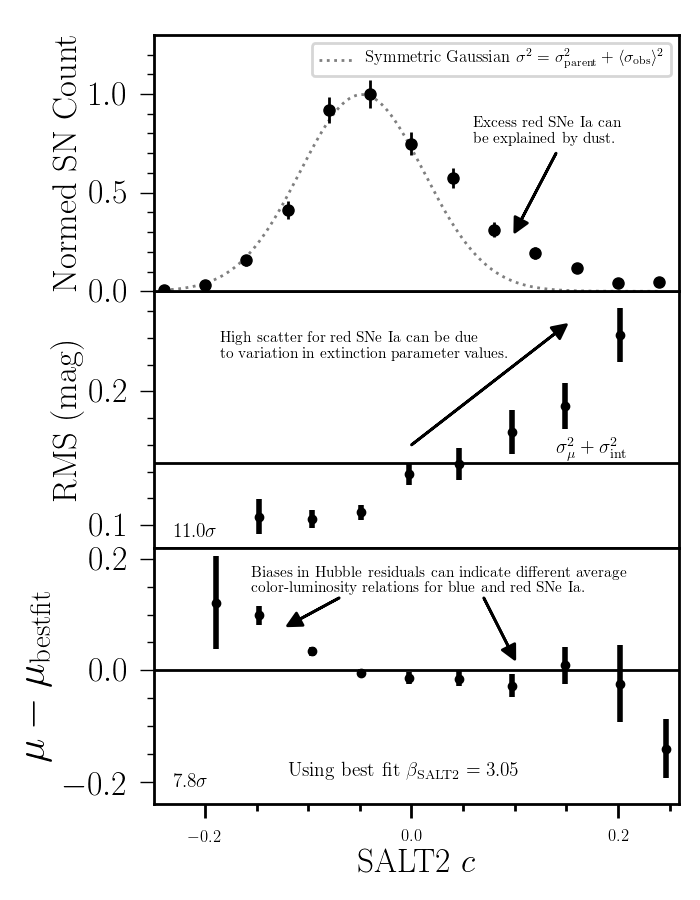}
\caption{\textbf{Top:} Observed color histogram from the full data sample, with symmetric Gaussian overlaid. \textbf{Middle:} RMS of Hubble Diagram residuals as a function of color. The RMS is calculated after Tripp standardization and after subtracting the mean Hubble residual bias. \textbf{Bottom:} Binned Hubble diagram residuals as a function of color, after Tripp standardization using the best fit $\beta_{SALT2}$. In the bottom two panels, the significance of the deviation from a flat line is show in the bottom corner. }

\vspace{.1in}
\label{fig:cartoonmetric}
\end{figure}

We define three metrics based on the three panels of Fig.~\ref{fig:cartoonmetric} which are pseudo $\chi^2$ evaluations that assess agreement between simulations that assume an SN~Ia model and the data. The first metric is defined as $\chi^2_{\rm c}$ for the agreement in color histograms of data  (${\rm N}_{\rm data_{c}}$) and survey simulations (${\rm N}_{\rm sim_{c}}$) such that
\begin{equation}
    \chi^2_{\rm c} = \sum_j ({\rm N}^{\rm data}_{c_j} - {\rm N}^{\rm sim}_{c_j})^2/e_{nj}^2,
\label{chisqc}
\end{equation}
and is determined in bins of color ($j$) where $e_{nj}$ is determined by Poisson statistics. 
    
A second metric, the agreement in total Hubble diagram scatter (RMS) between data (${\rm RMS}_{\rm data}$) and survey simulations (${\rm RMS}_{\rm sim}$), is defined as $\chi^2_{\rm RMS}$ over color bins
such that
\begin{equation}
    \chi^2_{\rm RMS} = \sum_i ({\rm RMS}^{\rm data}_{c_i} - {\rm RMS}^{\rm sim}_{c_i})^2/e_{ci}^2
\label{chisqrms}
\end{equation}
and is determined in bins of color $i$ and where $e_{ci}$ are the errors determined from 100 realizations of the simulated dataset. We use RMS instead of intrinsic scatter as a metric because, for intrinsic scatter, the sensitivity of the different components of the error modeling is difficult to track.

A third metric is the agreement in distance modulus residuals between data (${\rm \Delta \mu}^{\rm data}$) and survey simulations (${\rm \Delta \mu}^{\rm sim}$) which can be expressed as $\chi^2_{\rm \Delta \mu}$ over color bins
such that
\begin{equation}
    \chi^2_{\rm \Delta \mu} = \sum_i ({\rm \Delta \mu}^{\rm data}_{c_i} - {\rm \Delta \mu}^{\rm sim}_{c_i})^2/e_{\mu i}^2
\label{chisqdmu}
\end{equation}
and is determined in bins of color $i$ and where $e_{\mu i}$ are errors derived from the data itself.

 A fourth metric is the agreement between the recovered \citep{Marriner11} color-luminosity coefficients of simulations ($\beta_{SALT2}^{\rm sim}$) and data ($\beta_{SALT2}^{\rm data}$) such that

\begin{equation}
    \chi^2_{\beta_{SALT2}} = (\beta_{SALT2}^{\rm data} - \beta_{SALT2}^{\rm sim})^2 / (\sigma_{\beta_{SALT2}^{\rm data}}^2+\sigma_{\beta_{SALT2}^{\rm sim}}^2)
    \label{chisqbeta}
\end{equation}
where $\sigma_{\beta_{SALT2}}$ is the uncertainty reported following \cite{Marriner11}.

Finally, we minimize the cumulative $\chi^2$ in our fits:
\begin{equation}
    \chi^2_{\rm Tot} = \chi^2_{\rm c} + \chi^2_{\rm RMS} + \chi^2_{\rm \Delta \mu} + \chi^2_{\beta_{SALT2}} 
    \label{chisqtot}
\end{equation}

The fitting process involves large simulations which given current SNIa sim/analysis infrastructure is prohibitive for out-of-the-box minimizers and Monte Carlo samplers. We therefore implement the following minimization algorithm:
 \begin{enumerate}
 
\item Course grid minimization of model parameters for initial guess
\item Proposal of new model parameters
\item 100 simulations with 1-dimensional perturbations around proposed parameters (no covariance)
\item Gradient descent
\item Repeat steps 2-5 (typically around 50 iterations)
\end{enumerate}

Results using this algorithm and its limitations are discussed in Section 3.4.  We note that this method does not account for covariance between fitted parameters. This is the work of a future paper (Popovic et al in prep.) which incorporates significant infrastructure improvements.

\section{Evaluating Models of Type Ia Supernovae Colors and Intrinsic Scatter}
\label{sec:sims}
\subsection{Previous Models of Intrinsic Scatter and Associated Intrinsic Color Populations}
Recent studies have focused on two models of intrinsic scatter, which to first order, can both be described by two parameters: the magnitude of chromatic and achromatic scatter. The two models are the `G10' scatter model \citep{Guy2010} which prescribes 70\% of the intrinsic scatter to coherent variation and 30\% to chromatic (wavelength dependent) variation and the `C11' scatter model \citep{chotard11} which prescribes only 25\% of the intrinsic scatter to coherent variation but 75\% to chromatic variation. Both of these models were trained on data: for C11, it was trained on spectra from the SNFactory \citep{snfactory} and for G10, it was trained during the creation of the SALT2 model on a large subset of the light curves used in this analysis (\citealt{Guy2010}, B14).

These scatter models cannot be used in survey simulations to predict color distributions or the trends of Fig.~\ref{fig:cartoonmetric} without an associated color population and a $\beta_{SALT2}$ as defined in Eq.~\ref{eq:tripp}. For both the G10 and C11 scatter model, \cite{Scolnic16}, hereafter SK16, determined the underlying color population such that when it was combined with measurement noise, the color scatter from the scatter model, and selection effects, the observed color distribution matched that seen for the data in the top panel of Fig.~\ref{fig:cartoonmetric}. The underlying population was described by an asymmetric gaussian, with three free parameters. The value of $\beta_{SALT2}$ was determined by finding what input $\beta_{SALT2}$ in the simulations would yield an output $\beta_{SALT2}$ consistent with that found in the data from the methodology outlined in Section 2.2.  

The number of parameters that describe the framework for one of these scatter models is six: two parameters for the spectral and coherent scatter, three parameters for underlying population, and the value of $\beta_{SALT2}$. However, in order to explain inconsistencies between the low-z targeted sample and the high-z samples, SK16 determined the underlying population for each separately. Therefore, in total, a description of the full sample is described by 9 parameters. 

For the simulations with G10 and C11, a single input $\beta_{\rm SALT2}$ value is used for each one: $\beta_{\rm SALT2}=3.1$ and $\beta_{\rm SALT2}=3.8$ for G10 and C11 respectively. As explained in past analyses \citep{ScolnicMW,Scolnic16,BBC}, applying the 1D fitting procedure from \cite{Marriner11} recovers an observed $\beta_{\rm SALT2}\sim3.1$ for both the G10 and C11 cases. Due to the larger amount of color scatter in the C11 model, the associated underlying color population of C11 appears much more like a sharp dust-like exponential distribution \citep{ScolnicMW} than the one for G10. While it is unclear how to apply a physical interpretation to the G10+SK16 model, one possible interpretation for the C11+SK16 model is that there are two color-luminosity relations: one that relates the dust-like color to luminosity, and another with no luminosity correlation ($\beta=0$) for the intrinsic color distribution.  The populations used for the samples in Pantheon (Low-z, PS1, SNLS, SDSS) can be found in SK16, for Foundation in \cite{Jones18}, and for DES3YR in B19b.

\subsection{Evaluating Past SN~Ia Scatter Models}

As expected, because the SK16 populations were determined so that simulations would reproduce the observed color distribution of the data, simulations based on C11+SK16 and G10+SK16 show excellent agreement with the data (Figure \ref{fig:chist}): $\chi^2_c$ of 9.0 and 9.5 respectively (12 bins). The mean observed $c$ and $x_1$ for the simulations, binned over redshift, is shown in the bottom panels of Figure \ref{figone} and is in similarly good agreement with the data. However, the agreement between data and simulations for both the RMS (Fig. \ref{redscatter}a) and mean Hubble residuals (Fig. \ref{redscatter}b) is comparatively poor. 

For the RMS of Hubble residuals (Fig. \ref{redscatter}a), it is clear that neither G10+SK16 nor C11+SK16 produce the trend observed in the data. We do see non-linear behavior predicted from the simulations for the C11+SK16 model, which prescribes more scatter due to SN~Ia chromatic variation and achieves a $\chi^2_{\rm RMS}=35$, whereas G10+SK16, which prescribes little color variation, achieves a $\chi^2_{\rm RMS}=68$. The relatively flat dependence of the RMS on color as predicted from the G10+SK16 model shows that the trend in the data can not be explained by lower signal-to-noise for SNe with redder colors.

The agreement between data and simulations for mean Hubble residuals (Fig. \ref{redscatter}b) is somewhat better for  G10+SK16 ($\chi^2_{\rm \Delta \mu} \sim 12$) but worse for C11+SK16 ($\chi^2_{\rm \Delta \mu} \sim 29$). As discussed in SK16 and used for the motivation of \cite{BBC}, both models do predict the upturn in mean Hubble residuals for blue colors. Such distance modulus biases arise due to the combination of asymmetric color distributions with color scatter and selection effects.  

\begin{figure}
\centering 
\includegraphics[width=0.49\textwidth]{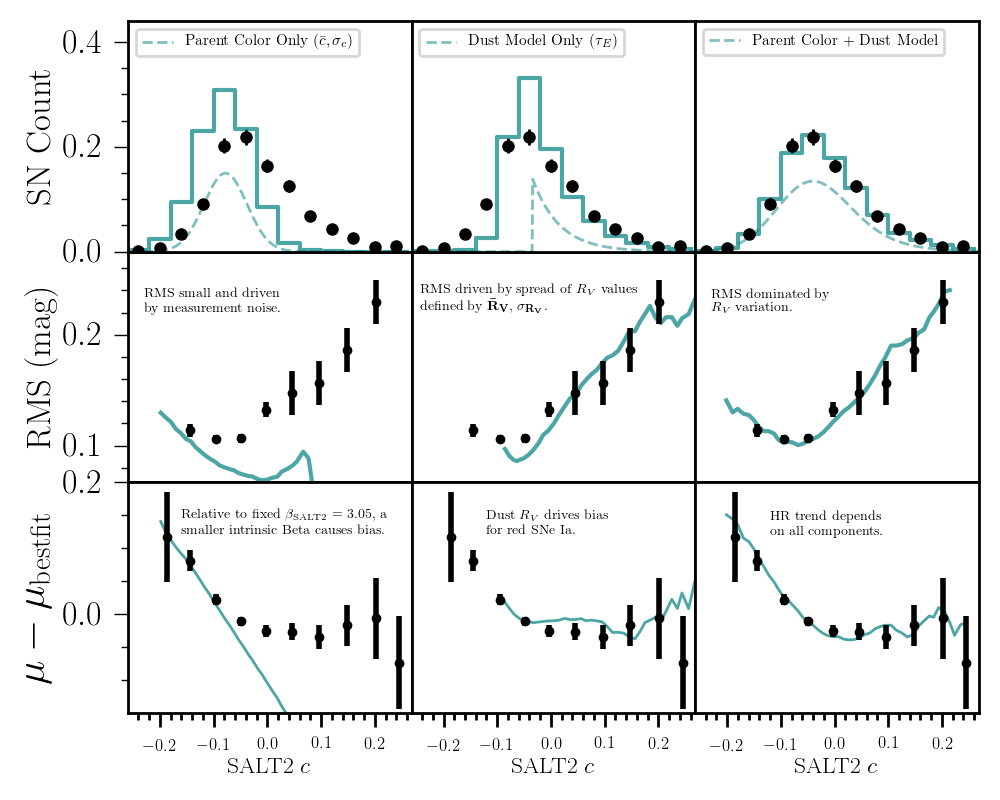}
\caption{Explaining the BS20 model. Shown are input parameterizations to simulations (dashed teal), the simulated values after measurement noise and selection effects (solid teal), and the dataset (black points) for the same three quantities (y-axes) as in Fig. \ref{fig:cartoonmetric}. \textbf{Left}: A simulation based solely on an intrinsic color distribution, described by a symmetric Gaussian,  without dust. \textbf{Middle}: A simulation based solely on a delta function in intrinsic color and an exponential dust distribution. \textbf{Right}: A simulation with both intrinsic color Gaussian and dust distribution combined.}
\label{fig:cartoon2}
\vspace{.07in}
\end{figure}


\subsection{Parameterization of a new dust-based color model}
\label{sec:model}

We present in Fig. \ref{fig:cartoonmetric} and Fig. \ref{fig:cartoon2} a simple and more physical understanding of the trends seen: the redder colors can be explained by dust extinction, the high RMS for red SNe~Ia could be explained by variations in the extinction parameter, and Hubble residual biases for the blue and red SNe can be explained by different respective color-luminosity relations. Here, we follow \cite{mandel11} and \cite{Mandel17}, which build on the work of \cite{Jha2007} to create a model of SN color based on two components: 1) an intrinsic color component ($c_{int}$) related to luminosity by a correlation coefficient $\beta_{\rm SN}$ and 2) a dust-component ($E_{\rm dust}$) described by an exponential distribution of reddening values related to luminosity by the extinction ratio $R_V$. The observed color $c_{\rm obs}$ can be expressed as
\begin{equation}
    c_{\rm obs} = c_{\rm int} + E_{\rm dust} + \epsilon_{\rm noise}.
\label{eq:cobs}
\end{equation}
where $\epsilon_{\rm noise}$ is measurement noise. We expand on the model from \cite{mandel11}  by allowing $R_V$ to be described by a Gaussian distribution to reflect that a range of values are seen in the literature, rather than a single value. In total, the model has seven fundamental parameters:
\begin{itemize}
  \item $\bar{c}$: the mean of the intrinsic color distribution described by a symmetric Gaussian.
  \item $\sigma_c$: the 1-sigma width of the  intrinsic color distribution described by a symmetric Gaussian.
  \item $\bar{\beta}_{\rm SN}$: the correlation between intrinsic color and luminosity.
    \item $\sigma_{\beta_{\rm SN}}$: the 1-sigma width of the Gaussian distribution from which the  correlation between intrinsic color and luminosity is drawn for each SN.
    \item $\bar{R}_V$: the center of the Gaussian distribution from which $R_V$ values are drawn for each SN.
  \item  $\sigma_{R_V}$: the 1-sigma width of the parent Gaussian $R_V$ distribution.
  \item $\tau_E$: the parameter describing the exponential distribution from which $E_{\rm dust}$ reddening values are drawn.
\end{itemize}

\begin{figure}
\centering
\includegraphics[width=0.46\textwidth]{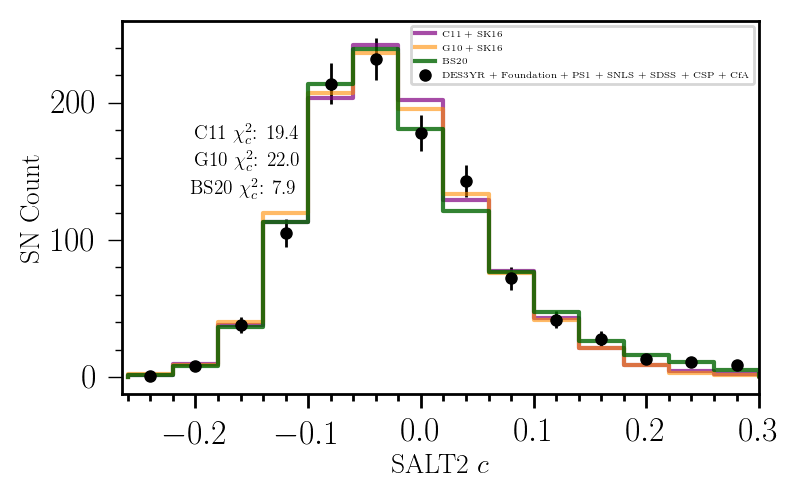}
\caption{A histogram of the observed color values from data (points) and simulations (lines). As all models are fitted so that simulations match the data for this metric, good agreement between data and simulation is expected for all the models.}
\label{fig:chist}
\end{figure}

\input{params}

\begin{figure*}
\begin{tabular}{cc}
  \includegraphics[width=0.49\textwidth]{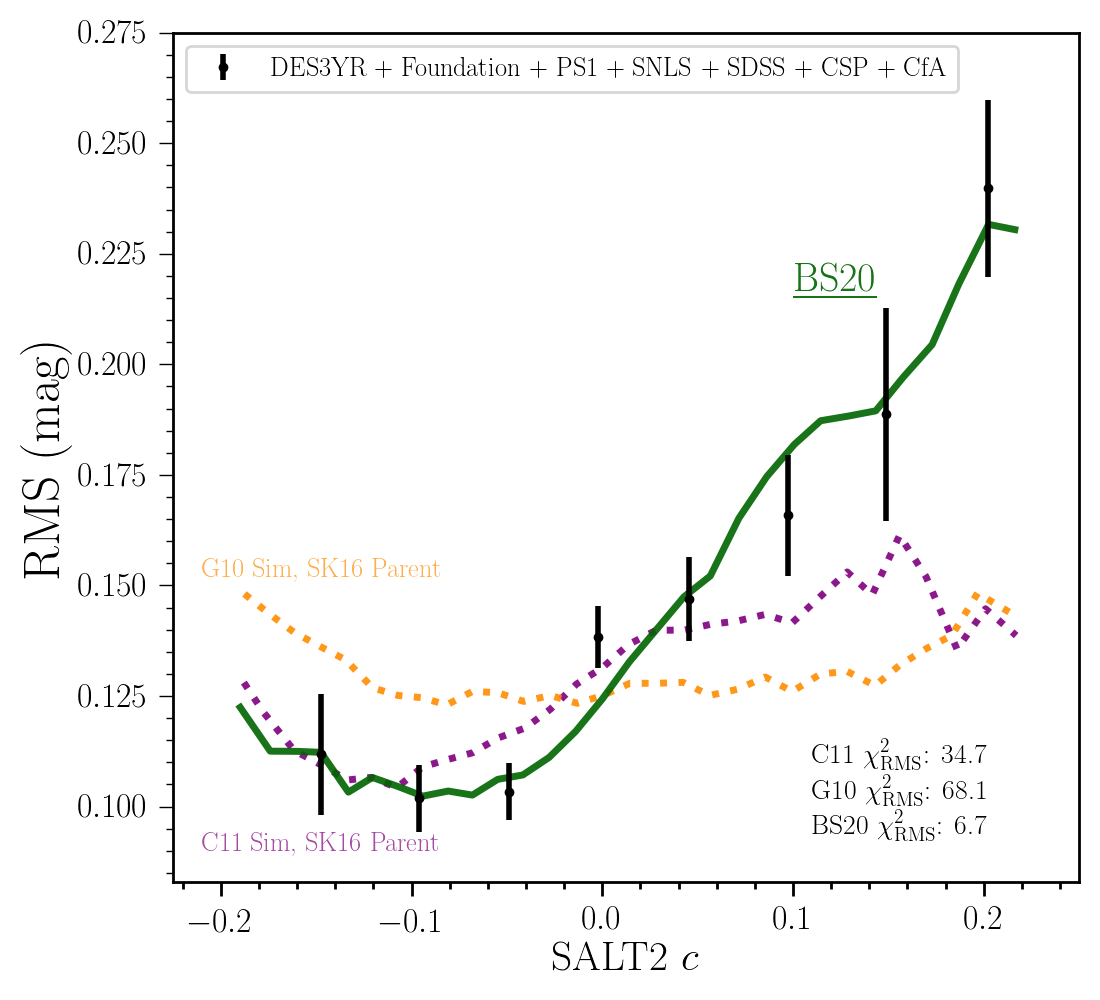} &   \includegraphics[width=0.49\textwidth]{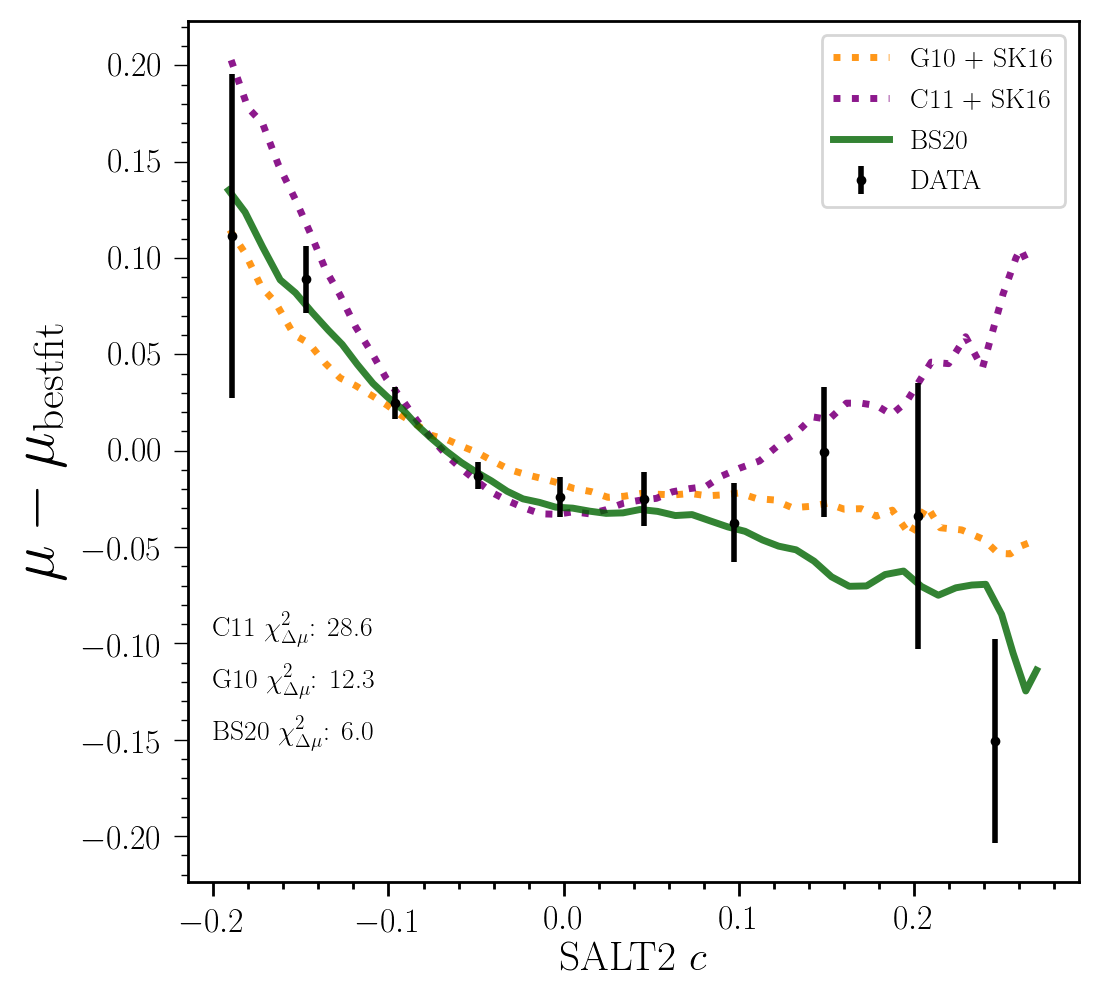} \\
\phantom{aaaaaaa}(a)  & \phantom{aaaaaaa}(b)  \\[6pt]

\end{tabular}
\caption{\textbf{a)} The zero-mean RMS of the Hubble residuals relative to $\Lambda$CDM versus the observed color $c$ of the SNe~Ia. The data is shown in black points, and the predictions from simulations of G10+SK16 and C11+SK16 models are shown in orange and purple dotted lines respectively. The model created for this work, labeled BS20, is shown in green. Inset: sames as main figure but for intrinsic scatter term $\sigma_{\rm int}$ instead of RMS. \textbf{b)} Binned Hubble Diagram residuals versus color. Biases are seen in the observed data (black points) and predicted by the scatter models (solid/dotted lines). For the BS20 model used here, there is no split on host-mass.}
\label{redscatter}
\end{figure*}

To set a `reddening-free' color, it is assumed that the intrinsic colors of SNe~Ia can be determined by:
\begin{equation}
    P(c_{\rm int}) =  \frac{1}{\sqrt{2\pi}\sigma_c}e^{-(c_{\rm int}-\bar{c})^2/2\sigma_c^2}.
\label{eq:colorprob}
\end{equation}
The reddening for each SN is described by $ E_{\rm dust}$ from Eq.~\ref{eq:cobs} and is related to the extinction of the SN by the standard equation
\begin{equation}
    A_V = R_V * E_{\rm dust}
\label{eq:avdustmodel}
\end{equation}
where $E_{\rm dust}$ corresponds to $E(B-V)$.

The reddening values $E_{\rm dust}$ are drawn from an exponential distribution following \cite{Mandel17} with probability density
\begin{equation}
    P(E_{\rm dust}) = 
    \begin{cases} 
      ~\tau_{E}^{-1}e^{-E_{\rm dust}/\tau_E} & ,~E_{\rm dust} > 0\\
      ~0 & ,~E_{\rm dust} \leq 0
   \end{cases}
\label{eq:dustprob}
\end{equation}
where $\tau_E$ is a parameter in the model described above.

In addition, we draw from distribution of possible values for $R_V$:
\begin{equation}
    P(R_V) = \frac{1}{\sqrt{2\pi}\sigma_{R_V}}e^{-(R_V-\bar{R}_V)^2/2\sigma^2_{R_V}} 
\label{eq:rvprob}
\end{equation}
where $\bar{R}_V$ is the center of the Gaussian distribution of $R_V$, $\sigma_{R_V}$ is the width, and where individual $R_V$ values below 0.5 are not allowed. 

Finally, similar to Eq.~\ref{eq:rvprob}, values for $\beta_{\rm SN}$ are drawn for each SN using model parameters $\bar{\beta}_{\rm SN}$ and $\sigma_{\beta_{\rm SN}}$ such that:
\begin{equation}
    P(\beta_{\rm SN}) = \frac{1}{\sqrt{2\pi}\sigma_{\beta_{\rm SN}}}e^{-(\beta_{\rm SN}-\bar{\beta}_{\rm SN})^2/2\sigma^2_{\beta_{\rm SN}}} .
\label{eq:bsnprob}
\end{equation}

In total, the change in observed peak brightness of a SN due to color can be expressed as $\Delta m_B$
\begin{equation}
    \Delta m_B = \beta_{\rm SN}c_{\rm int} + (R_V+1) E_{\rm dust} + \epsilon_{\rm noise}
\label{eq:betasn}
\end{equation}
where each observed parameter is unique to each SN. The coefficient $R_V+1$ is used rather than $R_V$ as in Eq.~\ref{eq:avdustmodel}, because to measure the change in $m_B$, the extinction parameter $R_B=R_V+1$ is needed.

To describe one survey with this model, seven parameters are required. If one is to solve for parameters to describe all high-redshift and low-redshift surveys separately, then one additional parameter is needed: a separate $\tau_E$ for each. In total, this makes 8 parameters. In contrast, as discussed previously, the G10+SK16 or C11+SK16 models require 9 parameters when high-redshift and low-redshift samples are accounted for separately. Thus, the dust-based framework described here has fewer free parameters than those used in past cosmological analyses.

\subsection{Results for the New Color Model}

The parameters described in Section \ref{sec:model} can be fit from the photometric data itself using the four metrics (Eqs. ~\ref{chisqc},~\ref{chisqrms},~\ref{chisqdmu},~\&~ \ref{chisqbeta}). Model parameters and their 1-d uncertainties are shown in Table \ref{tab:paramtable}.
We present the $\chi^2$ surfaces from our iterative forward-modeling minimization process in Figure \ref{fig:chigrid} and we show visually the degeneracies between model parameters in Appendix Fig. \ref{grid} (as well as for additional models). We note that the estimates of the uncertainties are limited by computational capability and thereby require coarseness of the model grid. While some of the posteriors are not clearly Gaussian, we assume Gaussianity to determine the uncertainty.  The only exception is when the posterior hits a cutoff (e.g. $\tau=0$) in which case we report upper and lower uncertainties.

We find a mean reddening-free color of $\bar{c}=-0.084\pm0.004$ with an intrinsic color distribution of $\sigma_c=0.042\pm0.002$ and a mean intrinsic color-luminosity correlation coefficient of $\bar{\beta}_{\rm SN}=1.98\pm0.18$. We find no evidence of $\beta_{\rm SN}$ variation (1.75$\sigma$ significance) with $\sigma_{\beta_{\rm SN}}=0.35\pm0.20$. The recovered ${\beta}_{\rm SN}$ is smaller than the traditional $\beta_{\rm SALT2}\sim3$ found when assuming a single correction for the full SN~Ia color and dust population simultaneously, and shows a relatively weak correlation between intrinsic color and luminosity in comparison to the contribution due to dust. We find the $R_V$ distribution for the dust component is best described by $\bar{R}_V=2.0\pm0.2$ and $\sigma_{R_V}=1.4\pm0.2$. The value of $\sigma_{R_V}=1.4$ indicates a wide range of $R_V$, though with a set-floor of $R_V=0.5$. 
Because a single color-luminosity relation is assumed in standardization, even though our simulations include a wide range of $R_V$ values, we find that the measured $R_V$ variation dominates the scatter of distance modulus residuals, contributing 0.095 to $\sigma_{\rm int}$, the majority of the total $\sigma_{\rm int}$ (0.106). On the other hand, the measured variation in the intrinsic color-luminosity relation ($\sigma_{\beta_{\rm SN}}=0.35$) contributes 0.040 to the total $\sigma_{\rm int}$ .

The results of simulations with our model are presented in Fig.~\ref{fig:chist} and Fig.~\ref{redscatter}. We find a $\beta_{\rm SALT2}=3.12\pm0.02$ when analyzed identically to the observed dataset, which is consistent with that of the observed dataset ($\beta_{\rm SALT2}=3.04\pm0.06$). In Fig.~\ref{fig:chist}, we show that the BS20 model results in observed SN~Ia color distribution similar to that of the data ($\chi^2_c\sim8$). Furthermore, as shown in Fig.~\ref{redscatter}a, this model captures the increased RMS scatter for the redder SNe ($\chi^2_{\rm RMS}\sim7$), which is attributed to variation of $R_V$. Finally, as shown in Fig.~\ref{redscatter}b, the BS20 model results in excellent agreement with observed Hubble residual biases ($\chi^2_{\rm \Delta \mu}\sim6$). 

In comparing $\chi^2$ values between the different color scatter models for the three metrics in Table~\ref{tab:chisq}, the advancement of the BS20 model is clear, and with one less parameter, the improvement cannot be simply attributed to additional model complexity. 

\input{chisq}

\begin{figure*}
\begin{tabular}{cc}
  \includegraphics[width=0.49\textwidth]{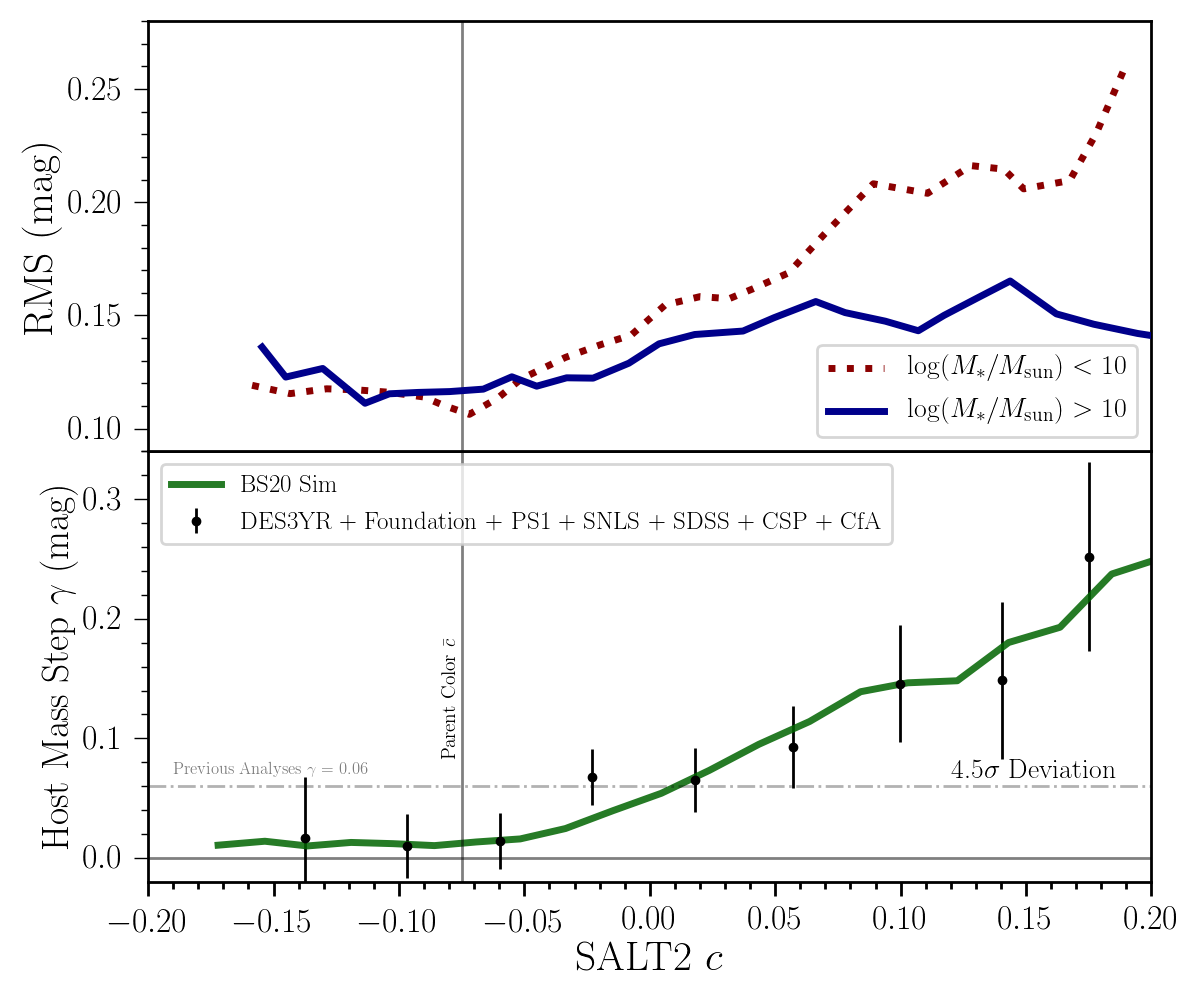} &   \includegraphics[width=0.49\textwidth]{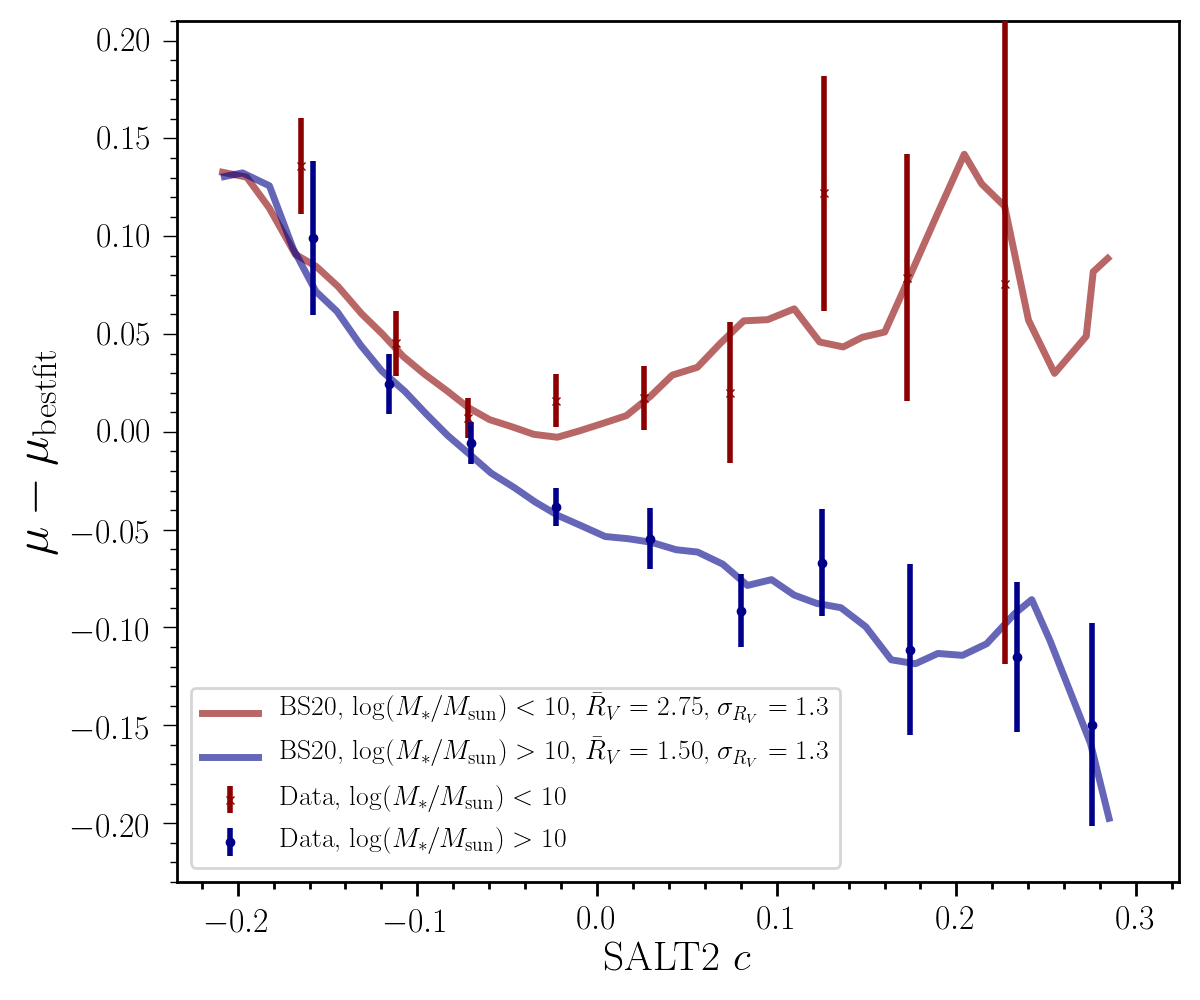} \\
\phantom{aaaaa}(a)  & \phantom{aaaaaaaaa}(b)  \\[6pt]

\end{tabular}
\caption{\textbf{a)} (Upper) Hubble Diagram scatter binned by SALT2 observed color and compared for SNe in host galaxies with low and high mass. 
(Lower) Recovered values of $\gamma$ for SNe Ia in high (log$(M_*/M_{\rm sun})>10$) versus low (log$(M_*/M_{\rm sun})<10$) mass hosts, binned by SALT2 observed color. Predictions from the BS20 Mass-split model is shown in green. Significance of the deviation from a constant $\gamma$ of 0.06 is shown (4.5$\sigma$). \textbf{b)} Binned Hubble Diagram residuals versus color split on host-mass. Biases are shown for the observed data (points) and predicted using the scatter models (solid lines). The difference between the red and blue points has typically been found by marginalizing over color and finding a single step $\gamma$. The dust parameters causing the observed split are shown in the legend.}
\label{massfig}
\end{figure*}

\section{The Dependence of the host-mass correlation with SNe Ia luminosity on color}

\subsection{Observed trends of color metrics based on Host-Galaxy Stellar Mass}
\label{sec:observed}

Many studies have found correlations between the the Hubble residuals and various host-galaxy properties \citep{Hicken09a,Lampeitl2010,Sullivan2010,Childress2013,Rigault2013,Roman18,Rose19}. Here, we focus on the host-galaxy stellar mass as it is the most commonly used, most accessible, and often yields some of the strongest correlations with Hubble residuals. In the top panel of Fig. \ref{massfig}a, the RMS versus SALT2 color plot as shown in Fig. \ref{redscatter}a is remade, but for the high and low host-mass subsamples separately.  For the `dust-free' blue SNe ($c\sim-0.1$), there is little difference between the RMS for SNe in low and high-mass hosts. However, the RMS increases with redder SN colors, and much more significantly for SNe in low-mass hosts.

As shown Fig.~\ref{massfig}b, when splitting the dataset into high and low host-mass subsamples, there is a distinct difference of the color dependence in the biases of Hubble residuals. For the `dust-free' blue SNe, the slope of the color-luminosity relation as well as the absolute Hubble residual biases for SNe in low and high-mass host subsamples are identical. For the redder SNe however, there are distinctly different color-luminosity relations and there is as much as a $\sim0.15$ mag difference in Hubble residuals. Overall, the subsamples are discrepant at greater than 5$\sigma$ ($\chi^2/N_{\rm bin}=57/10$) relative to each other.

Pursuing this further, we follow recent works like B19b and define $\gamma$ as the mean difference in Hubble residuals given a split in host galaxy properties:
\begin{equation}
\label{Eq:hm}
\delta\gamma = \gamma \times [1+e^{(\mathcal{M}_{\rm host}-\mathcal{M}_{\rm step})/0.01}]^{-1}-\frac{\gamma}{2},
\end{equation}
where $\mathcal{M}=\rm{log}(M_*/M_{\rm sun})$ and a log host-mass step location ($\mathcal{M}_{\rm step}$) of 10 is assumed. We determine $\gamma$ for the sample in discrete color bins. This is shown in the bottom panel of Fig.~\ref{massfig}a. As expected from the observations in Fig~\ref{massfig}b, for `dust-free' SNe~Ia that are bluer than the intrinsic color $\bar{c}$, $\gamma=0.003\pm0.029$, consistent with 0. However, for redder SNe, there is a significant $\gamma=0.083\pm0.011$ as well as a $4.5\sigma$ increasing trend where $\Delta \gamma \sim 0.72\pm0.14 \times c$, showing that the typical $\gamma$ values around 0.06 mag recovered in previous analyses are driven by the red SNe in the sample. 

 While many studies have shown that host-mass and SN color are weakly correlated if at all (e.g. \citealp{Sullivan2010}), the dependence of $\gamma$ itself on color has not been studied. As our model shows that redder colors can be described by dust, the difference between observed correlations between Hubble residuals and mass for different colors are all indicative of a dust-based explanation. We note that the trend seen in the bottom of Fig.~\ref{massfig}a is largely insensitive to whether distance bias corrections are applied. If we apply corrections based on \cite{BBC}, the $\gamma$ recovered is $0.0-0.02$ mag lower per bin than that shown, which is discussed at length in \cite{Smith2020}. The trend with RMS is not affected by these corrections because the RMS measured per bin is calculated after a mean offset is removed, thereby effectively doing a similar correction as \cite{BBC}.

\subsection{Dust Modeling Explains Mass Step}
\label{sec:hostmodeling}

We repeat the process as described in Section 3 for determining the underlying dust-based color model, except now for the low and high-mass host-galaxy subsamples separately. The fitted parameters are given in the `Mass-split' grouping of Table \ref{tab:paramtable}. Parameters that are intrinsic to the SNe~Ia are fixed for both host-galaxy subsamples while the dust distributions are allowed to vary for each subsample. We find that for SNe in low-mass hosts, $\bar{R}_V=2.75\pm0.35$ with $\sigma_{R_V}=1.3\pm0.2$, whereas for SNe in high-mass hosts, $\bar{R}_V=1.50\pm0.25$ with $\sigma_{R_V}=1.3\pm0.2$, suggesting that the peak $\bar{R}_V$ differ by 2.9$\sigma$ between low and high-mass hosts. We note that $\sigma_{R_V}$ is found to be the same between low and high-mass hosts, though it is unclear what the physical motivation for this would be. After accounting for selection effects, the distribution shifts such that the average observed $R_V$ for the detected SNe in the sample is 2.94 and 1.85 for low-mass hosts and high-mass hosts respectively. In these simulations, 2\% of all the detected SNe have simulated $R_V$ values greater than 5. The dust distribution for SNe in high-mass hosts that are discovered in high-$z$ surveys is described with $\tau_E=0.15\pm0.02$ whereas for low-mass hosts we find $\tau_E=0.12\pm0.02$ and similarly for the low-$z$ surveys the SNe can be described with $\tau_E=0.19\pm0.08$ whereas for low-mass hosts it is $\tau_E=0.01^{+0.05}_{-0.01}$. 

We show in Fig.~\ref{massfig}b that simulations with these separate dust models do indeed each recover the trends in Hubble residuals, and consequentially the trend seen in the bottom panel of Fig.~\ref{massfig}a. Therefore, we conclude that modeling different dust properties for different galaxy populations can fully explain the net $\gamma \sim0.06$ mag offset seen in past analyses as well as the $\gamma$ dependence on observed SN~Ia color.

As shown from the data, applying a single offset ($\gamma$) as has been done in past analyses, is incorrect. Furthermore, it has been unclear in past analyses why there should be any `step' behavior \citep{Sullivan2010}. Here, it is shown that the past step is an artifact of improper fitting, and arises because of significantly different $R_V$ distributions for different types of galaxies.


\section{Impact on Recovery of Cosmological Parameters}

\begin{figure}
\centering
\includegraphics[width=0.48\textwidth]{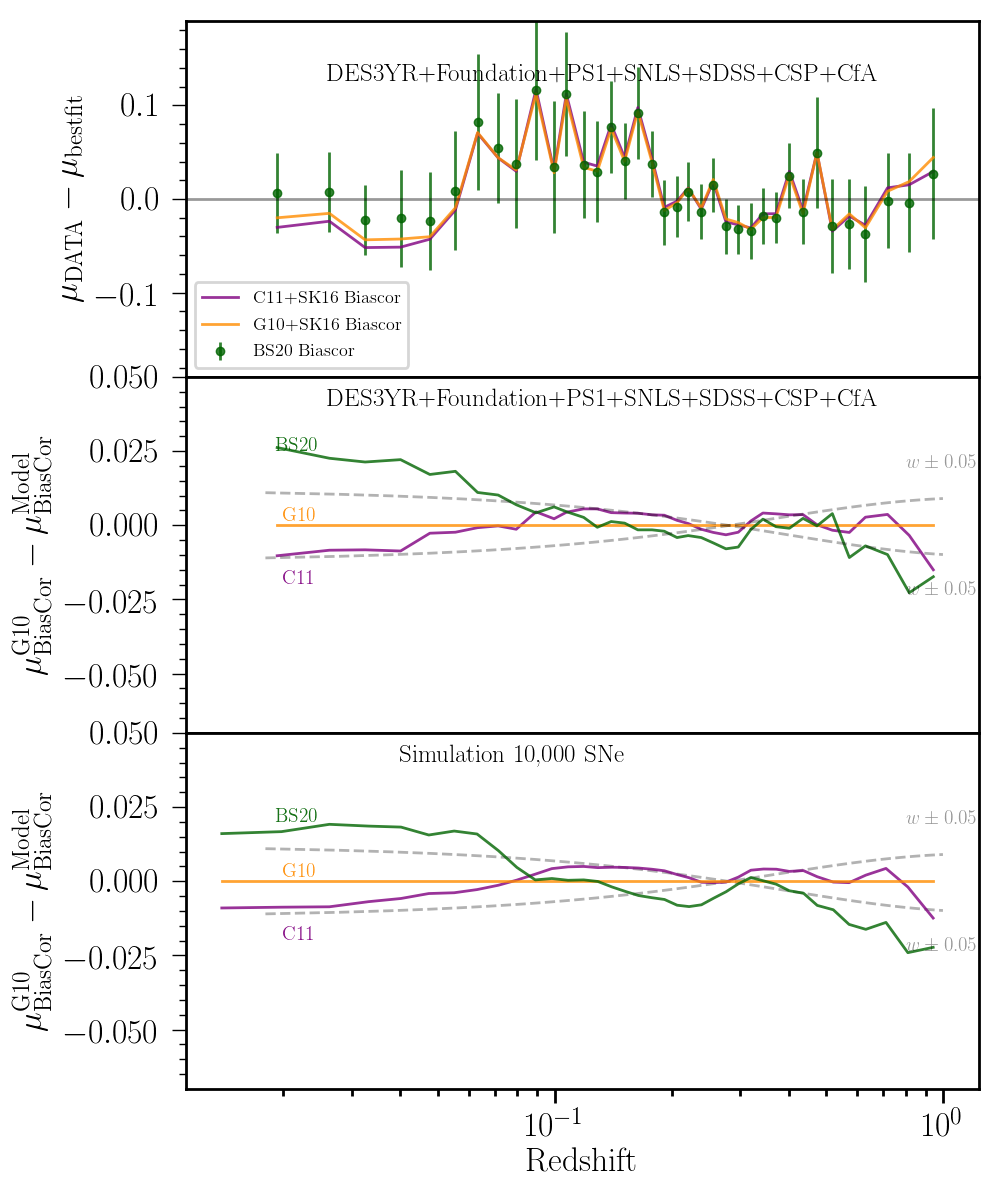}
\caption{
(Top) Hubble diagram residuals of the compiled DES3YR+Foundation+PS1+SNLS+SDSS+CSP+CfA dataset as a function of redshift. The dataset is bias corrected with the three different models of SN~Ia color. Error bars for G10+SK16 and `C11+SK16' are not shown and are indistinguishable from those of BS20. (Middle) The impact of bias corrections on real data relative to distances computed using G10+SK16.   (Bottom) The impact of bias corrections using simulated data relative to distances computed using G10+SK16.
}
\label{hdres}
\end{figure}

To understand the impact of these different models of SN~Ia color on the recovery of cosmological parameters, both data and simulations are used. Before measuring cosmological parameters, we apply bias corrections following the methodology of \cite{Marriner11} and B14 using large simulations with the three color models (G10+SK16, C11+SK16, BS20) to measure the dependence of distance biases with redshift, which are then applied as corrections to the dataset or a simulated dataset. Bias corrections following \cite{BBC} are not used because they have so far been only been designed to work given a $\beta_{\rm SALT2}$ and a variation in $c$, but not $R_V$, nor variation thereof. Therefore, we apply bias corrections that assume a single $\beta_{\rm SALT2}$ and follow the same formalism that was used in the JLA analysis and we do not split by host-mass. This is so a self-consistent comparison can be made against the impact of the G10+SK16 and C11+SK16 scatter models.

The impact of the bias corrections on the data is shown in Fig.~\ref{hdres}.  The most noticeable differences between the corrections of BS20 versus the other scatter models are at $z<0.1$ and $z>0.8$, where selection has the greatest influence. Here, the differences in recovered distance modulus can change by up to $\sim0.05$ mag at low or high-z depending on which color model is used. This difference is larger than any other systematic in past cosmology analyses (e.g., B19b). 

As shown in Fig. \ref{hdres}, we see the same effect with simulations as we do for data when simulating a sample of 10,000 SNe with realistic proportions and distributions of SNe Ia from each survey. Here, the simulations of `datasets' are based on the BS20 model, but bias corrections are determined from the other models.

 To determine cosmological parameters, we use CosmoMC \citep{cosmomc} and combine with CMB (Planck Collaboration et al. \citeyear{planck}) constraints. In Table \ref{wtable}, the biases in cosmological parameters are given when simulated SNe~Ia datasets use different models of SN~Ia color than the model used for the determination of distance bias correction. We find that if the `true' model of SN~Ia color is the dust-based model presented in Section \ref{sec:model}, but the bias corrections are based on the G10+SK16 or C11+SK16 models, the propagated bias in $w$ will be -0.025 and -0.040 respectively. Again, this bias is larger than any other systematic uncertainty reported in recent cosmological analyses. 

In Table \ref{wtable} we also show the differences in $w$ for the real data when we apply bias corrections based on simulations using the three separate models of color: G10+SK16, C11+SK16 and BS20. Relative to BS20 bias corrections, there are changes in recovered $w$ for G10+SK16 and C11+SK16 of -0.033 and -0.041 respectively, 
which is consistent with simulations. Interestingly, as shown in Fig. \ref{redscatter}a, while C11 and BS20 better match the trend in the data, they produce the largest differences in $w$ of $\sim$0.04.

\input{wtable.tex}

\section{Discussion} 
\label{conclusion}

\subsection{The Dependence Between $R_V$ and Host Galaxy Properties}

That the mass correlation can be explained by separate dust properties is now the only \textit{direct} explanation for the correlation between host-mass and distance modulus residuals. This possibility was briefly discussed in \cite{Mandel17}, which showed if one changed the dust distribution ($\tau_E$) for the SNe in low and high-mass subsamples, one could remove $1/3$ of the magnitude of $\gamma$, but not the whole effect. We follow this idea from \cite{Mandel17}, but add that the $R_V$ distribution as well should be different for these subsamples. This can then explain the full $\gamma$ as well as its color dependence. Our dust explanation aligns well with the observations in \cite{Burns18} that at low-z, the host-mass correlation with SN~Ia luminosity is larger in the optical than in the NIR, where the correlation is consistent with 0. This should be the case if the correlation is tied to reddening, as the corresponding extinction ratio of $R_V$ in the NIR is smaller.  Furthermore, the range of $R_V$ values is in good agreement with studies of $R_V$ from individual SNe like in \cite{Amanullah15}. While the model shows that a fraction of SNe should have $R_V$ above 5, we find that this is only $6\%$ after accounting for selection effects.

This analysis makes a strong prediction that SNe in lower-mass galaxies have on average, higher $R_V$ values than SNe in higher-mass galaxies. As there are very few measurements of $R_V$ in the interstellar medium of galaxies beyond the Milky Way, LMC and SMC, it is difficult to find evidence that this trend would hold for galaxies themselves. \cite{Salim18}, which measured the dust attenuation curves of 230,000 individual galaxies in the local universe, found that quiescent galaxies, which are typically high-mass, have a mean $R_V=2.61$ and star-forming galaxies, which are lower-mass on average, have a mean $R_V=3.15$. This trend is in general agreement with our prediction. 

The observation that global properties of the galaxy can impact the dust measured from the SNe is supported by \cite{Phillips13} and \cite{Bulla18}, which found the dust responsible for the observed reddening of SNe Ia appears to
be predominantly located in the interstellar medium of the host galaxies and not in the circumstellar medium associated with
the progenitor system. It's also supported by \cite{Childress_2013} which showed that color of SNe Ia is strongly tied to the metallicity of the host galaxy. For a future analysis, it is encouraged to repeat this same exercise but instead of using stellar mass to use metallicity, specific star formation rate, or local color; improved estimates of the dust distribution parameters would likely be obtained. For example, as shown in \cite{Sullivan2010}, when measuring a single color-luminosity coefficient $\beta_{\rm SALT2}$ for different samples, there is an even bigger difference when splitting the sample for specific star formation rate than there is for mass. As our model constrains both the amount of dust and the properties of dust itself, it is likely that different galaxy properties (e.g., distance to host and inclination, \citealp{Holwerda15,Galbany12}) will yield complementary insight about both of these components. We stress that our analysis does not limit the use of host-galaxy information in cosmological studies with SNe Ia, but rather, proposes a new path forward.

Indirect explanations of $\gamma$ have suggested that SNe from different progenitor systems have different luminosities, and the progenitor system can be potentially linked to the age of the host galaxy \citep{Childress2013}. However, any model that assumes that the luminosity depends on progenitors does not predict the key observation in our analysis that the magnitude of $\gamma$ depends on color. A progenitor-based explanation has motivated studies by \cite{Rigault2013}, \cite{Childress14}, \cite{Jones2015}, \cite{Jones18global}, and \cite{Roman18}, which focus on the local specific star formation rate, local mass, and local color. Some of these studies seem to indicate that measuring the local color produces the highest correlation with measured SN luminosity. In light of our dust-based SN~Ia color model, a simple explanation is that the local host color yields insight about the amount of dust and/or dust properties at the position of the SN. Our model  does not differentiate whether the dust is in the circumstellar surrounding which is still linked to the progenitor or in the interstellar medium which is not linked to the progenitor, but we can rule out a luminosity dependence on the progenitor system.  

Relatedly, many studies have found correlations between spectral features and Hubble residuals \citep{Fakhouri15,Siebert20}. Interestingly, \cite{Wang09} split a sample of 158 SNe Ia based on whether their spectra indicate  `normal velocity' or `high velocity' features, and find $R_V = 2.36 \pm 0.07$ and $1.57\pm0.07$ for the two subsamples respectively. \cite{Pan15} show that the velocity of spectral features correlates with the mass of the host galaxies, such that high-mass host galaxies regularly have high-velocity SNe, so one would expect low $R_V$ to be found for high-mass hosts. This is in great agreement with the results of our study, though we note that \cite{Foley11} show that different $R_V$ from \cite{Wang09} depend on using SNe with very red colors $E(B-V)>0.4$. As velocity features have typically been thought of indicative of properties of the progenitor and circumstellar surrounding, it is unclear at what level this is causally connected versus correlated.  

 As discussed in the introduction, circumstellar dust surrounding SNeIa has been used to explain low values of $R_V$ ($<2$) (e.g., \citealp{Goobar08}).  Circumstellar dust has also been used to explain similarly low $R_V$ values found for core-collapse SNe  \citep{Nugent2006,Goobar08}. This is supported by our findings that it is common for SNeIa to have $R_V<2$, though as part of a larger range from $R_V$ from 0.5 to 8. The interplay between SN radiation and nearby dust, and what can be learned by detailed studies of studies of individual SNe of different types, will be an important avenue to support or refute this possible explanation.

\begin{figure}
\centering
\includegraphics[width=0.48\textwidth]{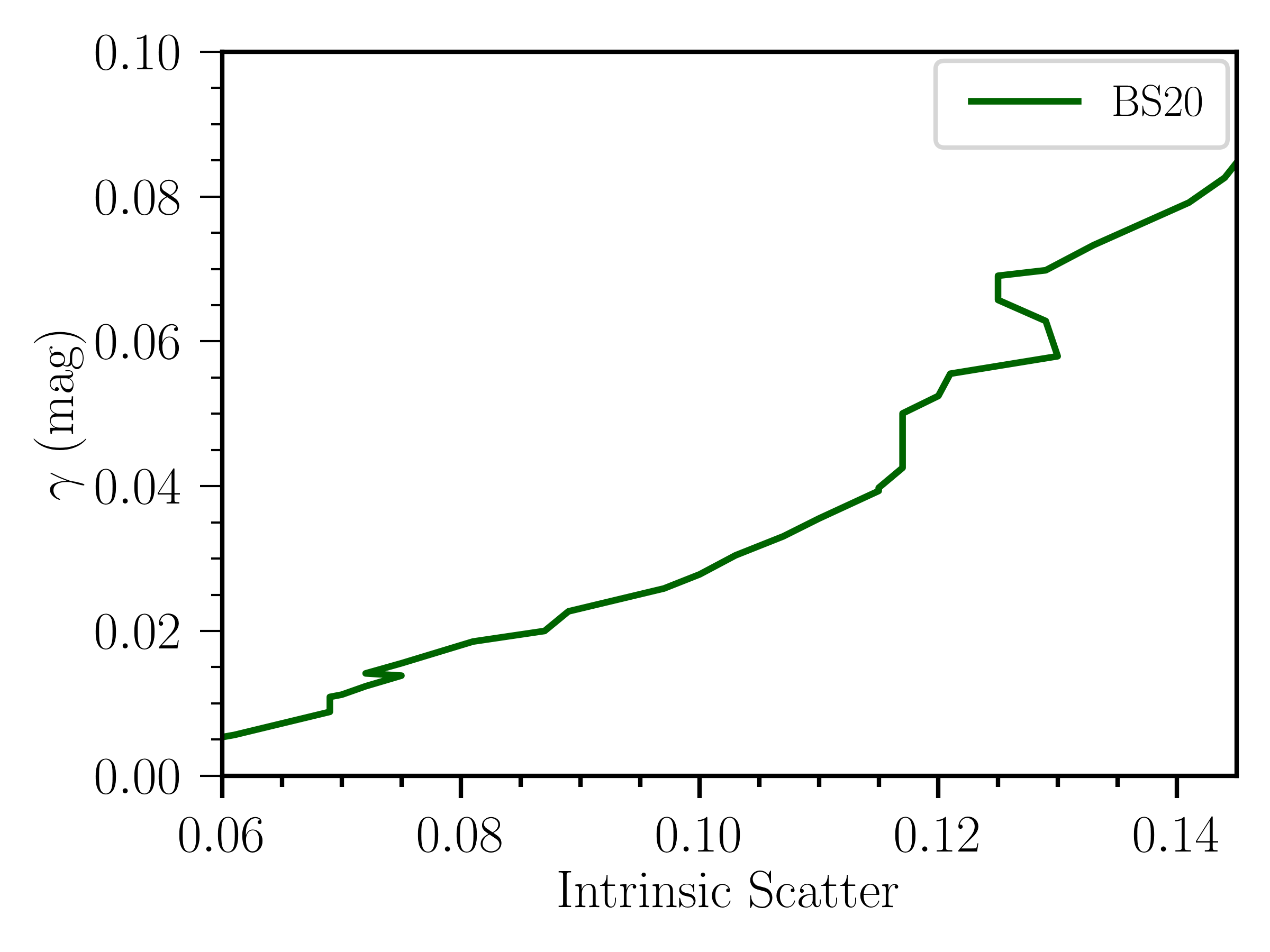}
\caption{  Simulations with the BS20 dust-based model predict a correlation between $\gamma$ and observed intrinsic scatter. This correlation was originally seen for real data in B19b.}
\label{fig:gvsc}
\end{figure}

\subsection{Application of BS20 Model In Future Analyses}

While we have shown that biases in $w$ from our model relative to previous models would have been the largest systematic uncertainty of previous analyses, there is a clear path to utilize this model for future analyses. In order to do so optimally, there are three necessary improvements. First, a full Bayesian fit to solve for the intrinsic and extrinsic parameters, broken by survey, redshift range, or targeted versus un-targeted is needed. This could be facilitated by the recent advancements by \texttt{Pippin} \citep{Pippin}. Future work can fully constrain and characterize systematic uncertainties on all 9 dust and hyper-parameters using a combination of the $\chi^2$ metrics. Second, this model should be integrated into the SALT2 training, which currently only accounts for one component of the observed SN~Ia color. Evidence of the benefit of retraining is shown in the Appendix Fig.~\ref{fig:restframe}. It will be necessary in the future to attempt to train SALT2 based on intrinsic and extrinsic color components. 

Third, as discussed in Section 5, BBC5D from \cite{BBC} is not capable of bias correcting two effective color-correlation coefficients ($\beta_{\rm SN}$ \& $R_V$). Additionally future analysis should not be correcting for an observed $\gamma$. Rather, dust distributions should be fit to different subsamples of host galaxies and using this information, distance bias corrections can be computed as a function of observables ($c$, $x_1$, $z$ and host galaxy properties). A future approach to such bias corrections (Popovic et al. in prep.) would be similar to that of \cite{BBC}. If done properly, we predict that there will be no residual $\gamma$ in the distance modulus residuals. Doing so will also improve the comparison of cosmological constraints in Section 5, where we had to assume naive mass-independent bias corrections. Ultimately one should compare the impact of the bias corrections from the two-mass model to the bias corrections from the G10 and C11 model when a luminosity-correction due to host-mass is applied. 

The difference in RMS for `dust-free' blue colors (RMS$\sim$0.1) versus redder colors (RMS$\sim0.18$) is striking. The statistical weight of these different SNe~Ia when constraining dark energy with our improved color model shows that that a blue SN~Ia is $\sim3\times$ more constraining than a red SN. The blue SNe~Ia exhibit an RMS at the same level as NIR SN~Ia standardized luminosities \citep{mandel11}. With tighter color measurement cuts and state of the art samples (SNLS, DES), we have seen that the `dust-free' RMS can even be as low as 0.08. In addition, as shown in \cite{BBC}, bias corrections are much smaller for blue SNe~Ia than red SNe~Ia. Additionally, because $\gamma$ is found to be consistent with 0 for the blue SNe, it appears that there are numerous advantages to using a sample of solely blue SNe. 
As LSST \citep{Ivezic} and WFIRST \citep{Spergel15,Hounsell18} will discover thousands of SNe~Ia in this un-extincted regime ($c\sim-0.1$), the vast difference in constraining power and intrinsic scatter for the blue SNe~Ia compared to the red SNe~Ia should be considered in planning survey strategy.

B19b showed an interesting trend that the magnitude of the recovered intrinsic scatter from various SN samples is correlated with the recovered $\gamma$ from that sample. They also remarked that the $\sigma_{\rm int}$ value of the low-z sample was more than 3$\sigma$ discrepant from that of the DES3YR sample. However, we show in Fig.~\ref{fig:gvsc} that this behaviour arises naturally from the BS20 model: both $\sigma_{\rm int}$ values are indeed consistent with a dust-based interpretation and that the relation between recovered $\sigma_{\rm int}$ and $\gamma$ is a direct prediction of the BS20 model. This is because the different distributions of observed colors for each sample imply different amounts of dust, different amounts of intrinsic scatter, as well as different magnitudes of $\gamma$.

With our new model, we showed that that the bias in recovered $w$ due to assuming the incorrect scatter model is $\sim$0.04, larger than any other systematic uncertainty quantified in recent SN~Ia cosmology analyses. In \cite{Brout18-SYS}, the systematic uncertainty ascribed to this issue was determined from averaging distance modulus values after applying bias corrections based on both the G10+SK16 and C11+SK16 model, thus halving the difference between them, but still found it to be one of the largest at $\sigma_w=0.017$. As the sensitivity of cosmological parameters to different scatter models is so large, we emphasize that this issue cannot be ignored in any future cosmological analysis. This statement is true for analyses of $w$ and for analyses of $H_0$ as well.
\cite{dhawan20H0} estimates biases due to scatter models to be on the level of $0.5-1.0\%$ in $H_0$. As the $H_0$ measurement has different systematic sensitivity than $w$ due to the comparison of SNe in calibrator galaxies versus Hubble flow galaxies, we recommend these two samples to have similar demographics of blue and red SNe. A full systematics treatment, as done in \cite{dhawan20H0}, should be done using the new dust-based SN~Ia color model described in this paper. Furthermore, we note that past discussions (e.g., \citealp{Rigault2013,Jones18global}) about potential biases in $H_0$ should be reconsidered in light of this paper's findings.


\section{Conclusion}
In this paper, we introduced a new, physical, two-component color model of SNe~Ia with an intrinsic component modeled as a simple symmetric Gaussian that correlates with SN~Ia luminosity and an extrinsic component that can be modeled by a dust distribution that is tied to extinction by a wide $R_V$ distribution. This model has fewer free parameters than previous models of SN~Ia color and a more physical motivation that better matches the data. Our findings suggest that the dominant component of observed SN~Ia intrinsic scatter is from $R_V$ variation of the dust around the SN. We also show that there is a \gsigma dependence on color of the correlation of host-mass with distance modulus residuals. Strikingly, this shows that previously observed host-galaxy property correlations with SN~Ia luminosity are driven by the redder SNe of the sample. This also suggests a dust-based explanation for the host-galaxy property correlations. By allowing our model to have different parameters for the dust distributions of SNe in high-mass versus low-mass host-galaxies, we show that the correlation between distance modulus residuals and host-galaxy stellar mass can be attributed to a 2.9$\sigma$ difference in $\bar{R}_V$ between low and high-mass.

By finding that the previously seen host-galaxy correlation with SN~Ia luminosity after standardization is actually due to differences of dust, and not due to possible variation in the luminosity based on progenitor systems, we find that that there is a tremendous amount of leverage to continue to improve cosmological analyses by studies of larger samples, measurements covering larger wavelength ranges, more host galaxy properties examined and improved dust models. Our study shows that so many disparate analyses of SNe Ia are actually intricately connected, and unifying these studies will provide tremendous improvements to measurements of the expansion of the universe.

\section{Acknowledgements}
We thank Rick Kessler, Adam Riess, Saurabh Jha, The Goobar Research Group, David Jones, Mat Smith, Doug Finkbeiner, Eddie Schlafly, Charlie Conroy, Antonella Palmese, and Sam Hinton for very useful discussions. We are appreciative of Rick Kessler for his ever-useful \texttt{SNANA} package. DB acknowledges support for this work was provided by NASA through the NASA Hubble Fellowship grant HST-HF2-51430.001 awarded by the Space Telescope Science Institute, which is operated by Association of Universities for Research in Astronomy, Inc., for NASA, under contract NAS5-26555. DS is supported by DOE grant DE-SC0010007 and the David and Lucile Packard Foundation. DS is supported in part by NASA under Contract No. NNG17PX03C issued through the WFIRST Science Investigation Teams Programme.

\appendix

\begin{figure*}
\begin{tabular}{cccc}
  \includegraphics[width=0.23\textwidth]{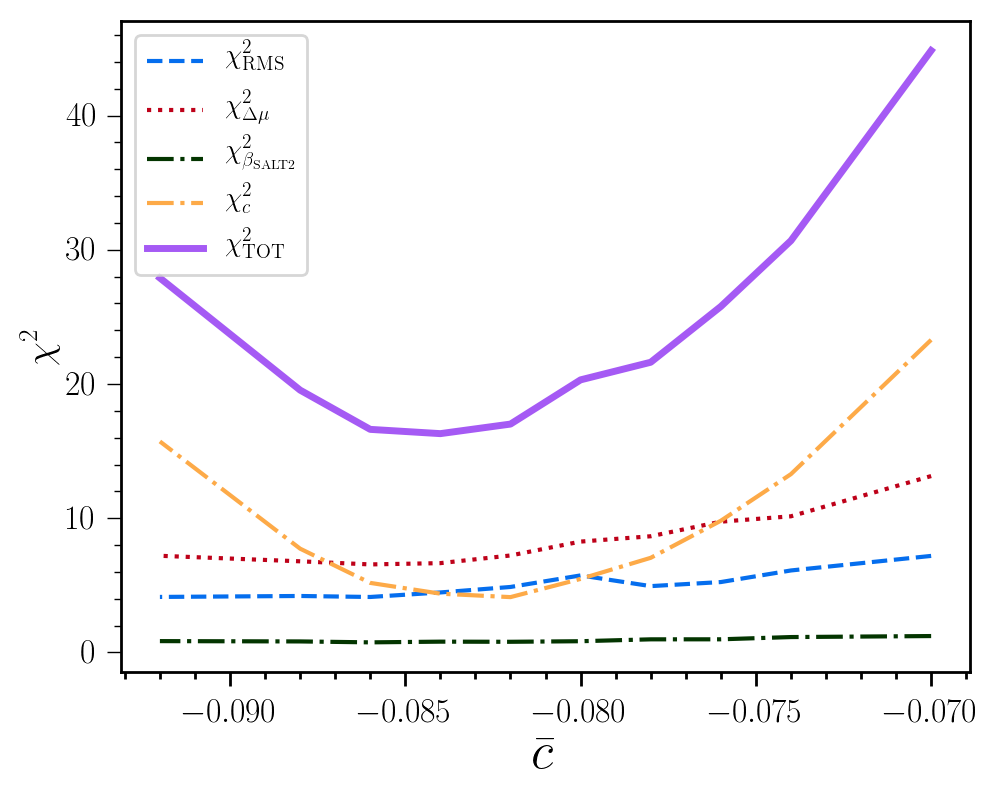} &   
    \includegraphics[width=0.23\textwidth]{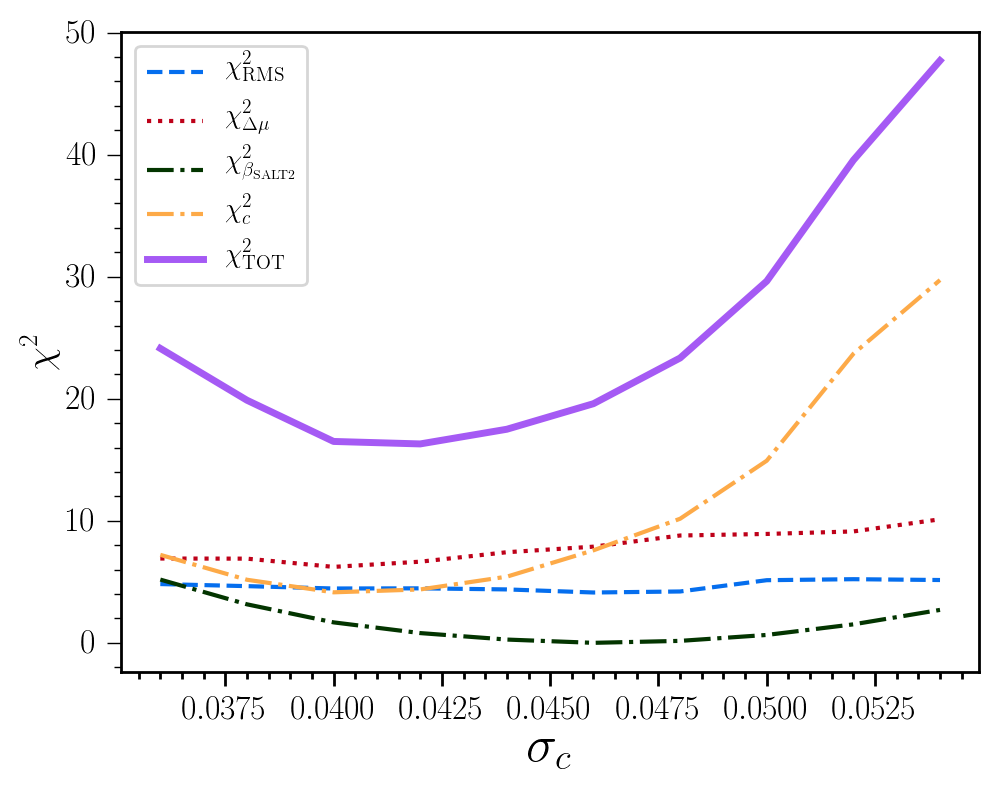} &  \includegraphics[width=0.23\textwidth]{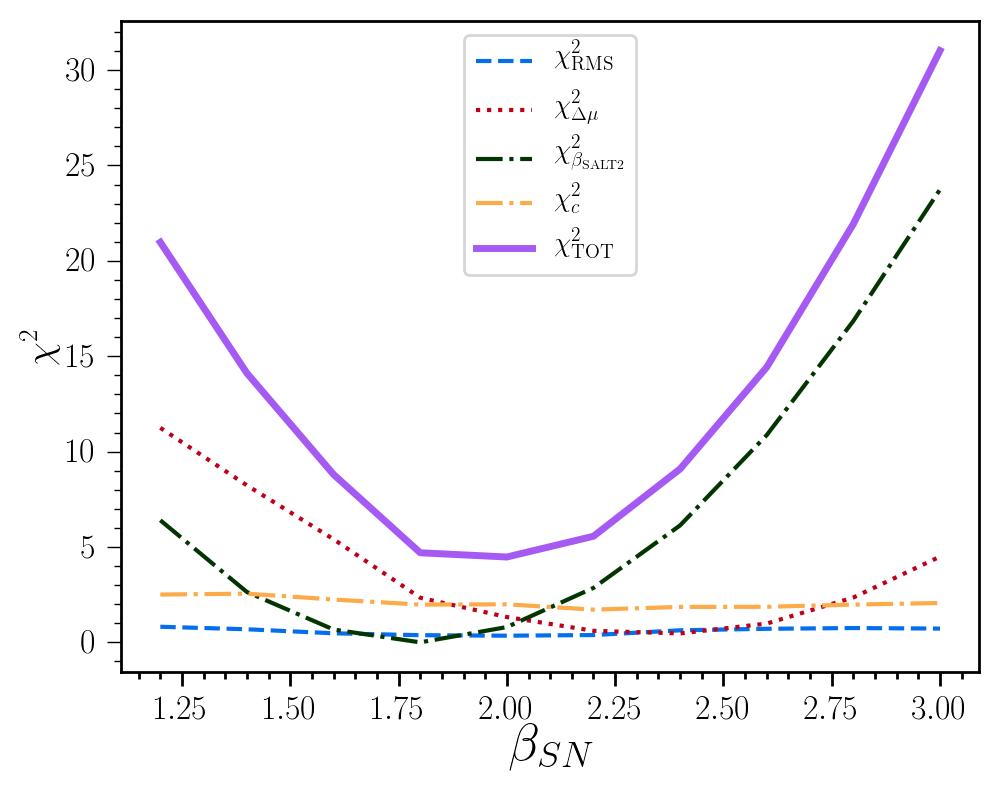} &    \includegraphics[width=0.23\textwidth]{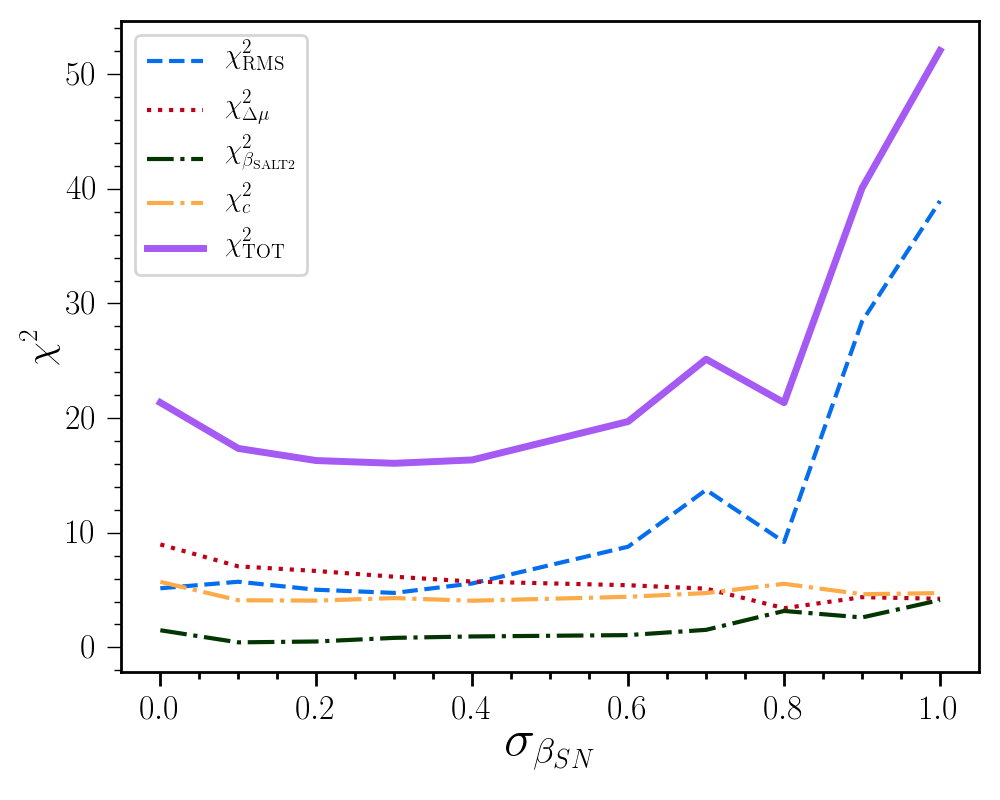} \\
  \includegraphics[width=0.23\textwidth]{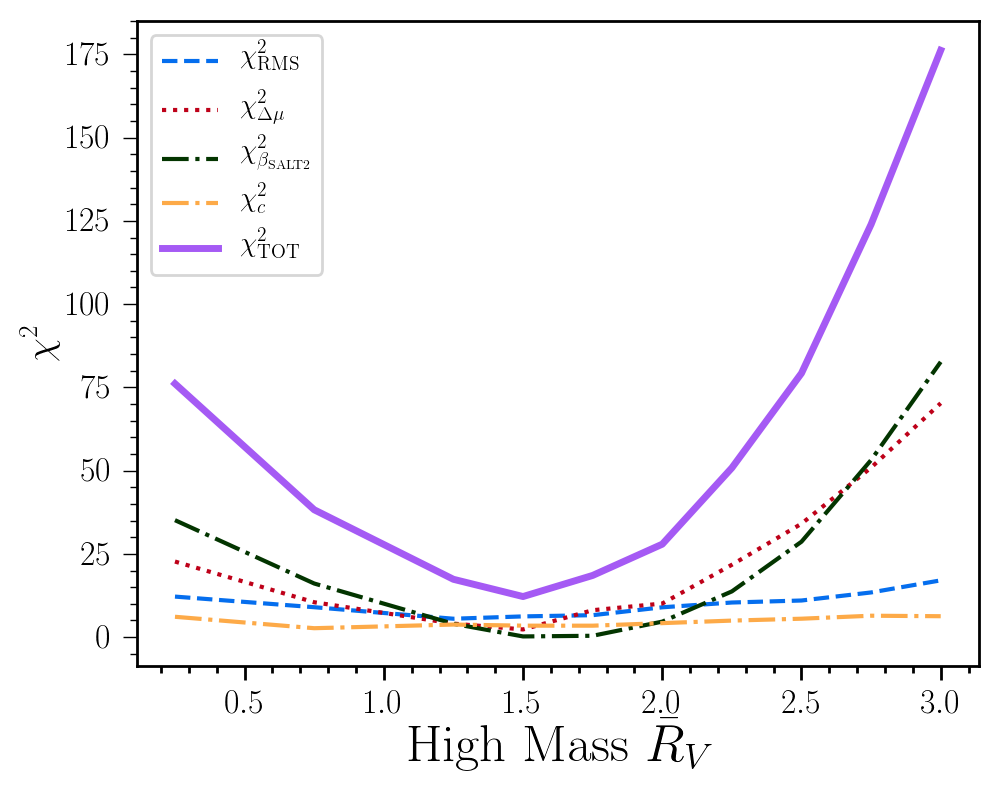} &   \includegraphics[width=0.23\textwidth]{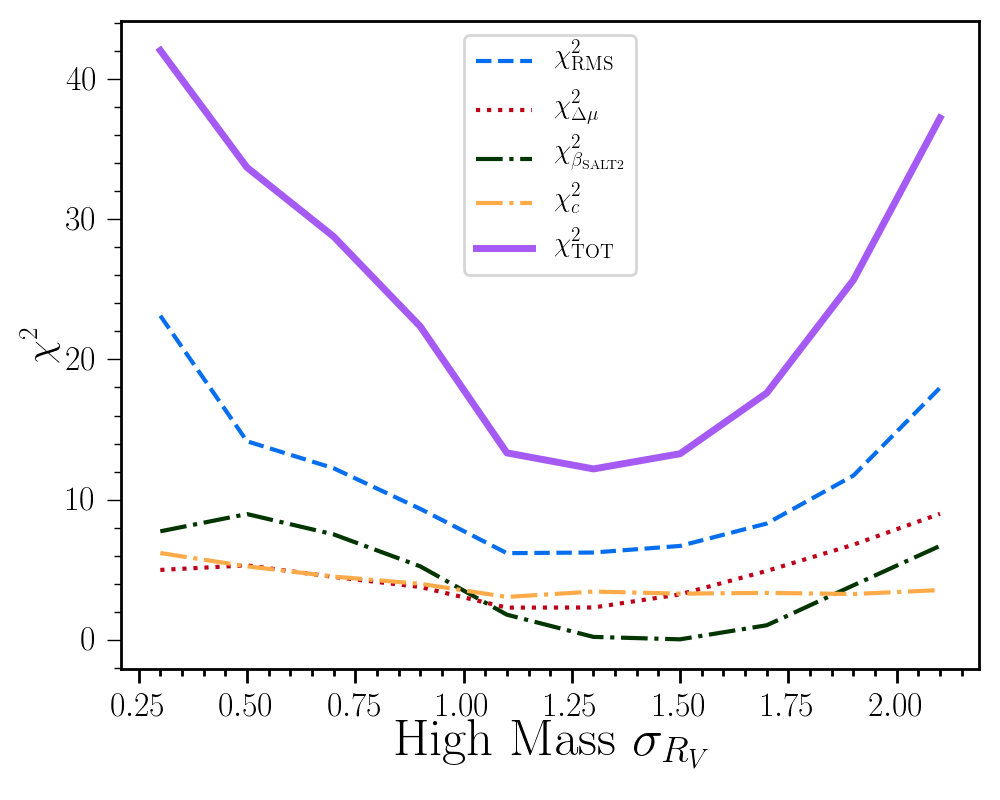} &
  \includegraphics[width=0.23\textwidth]{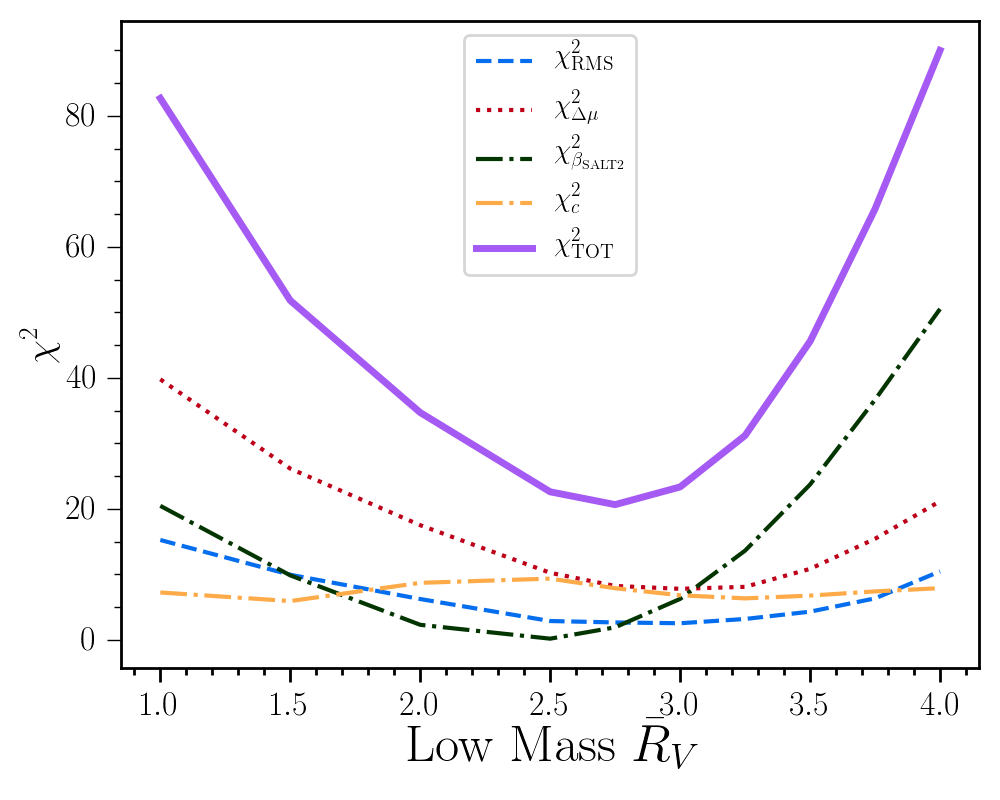} &   \includegraphics[width=0.23\textwidth]{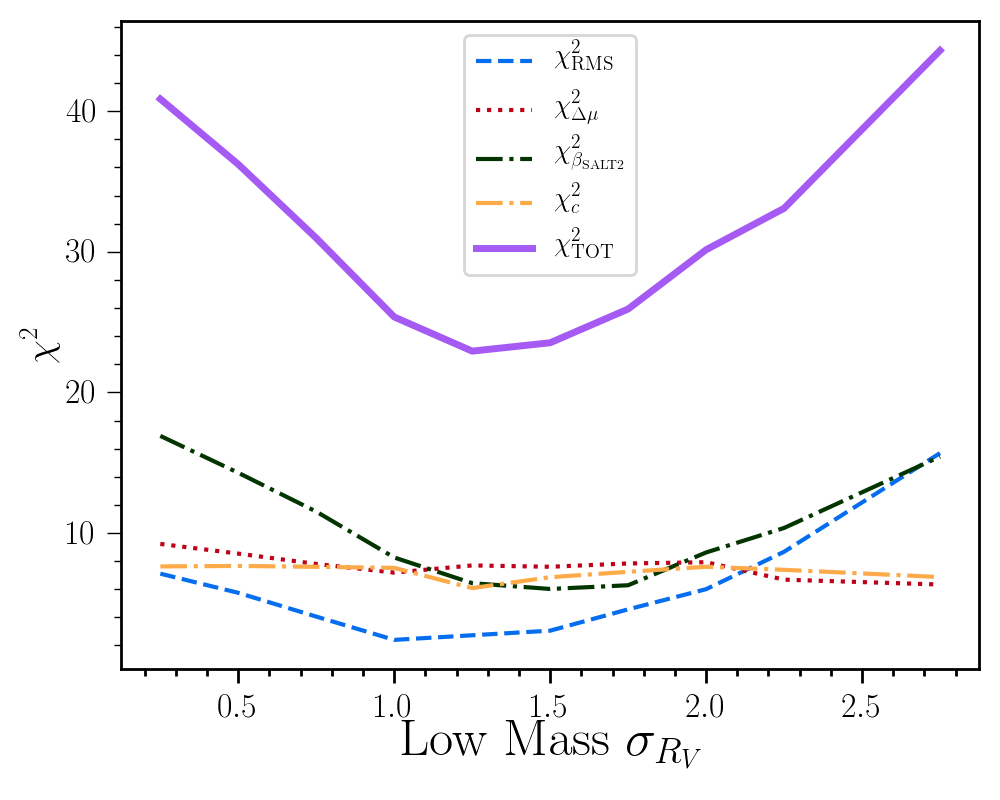} \\
 \includegraphics[width=0.23\textwidth]{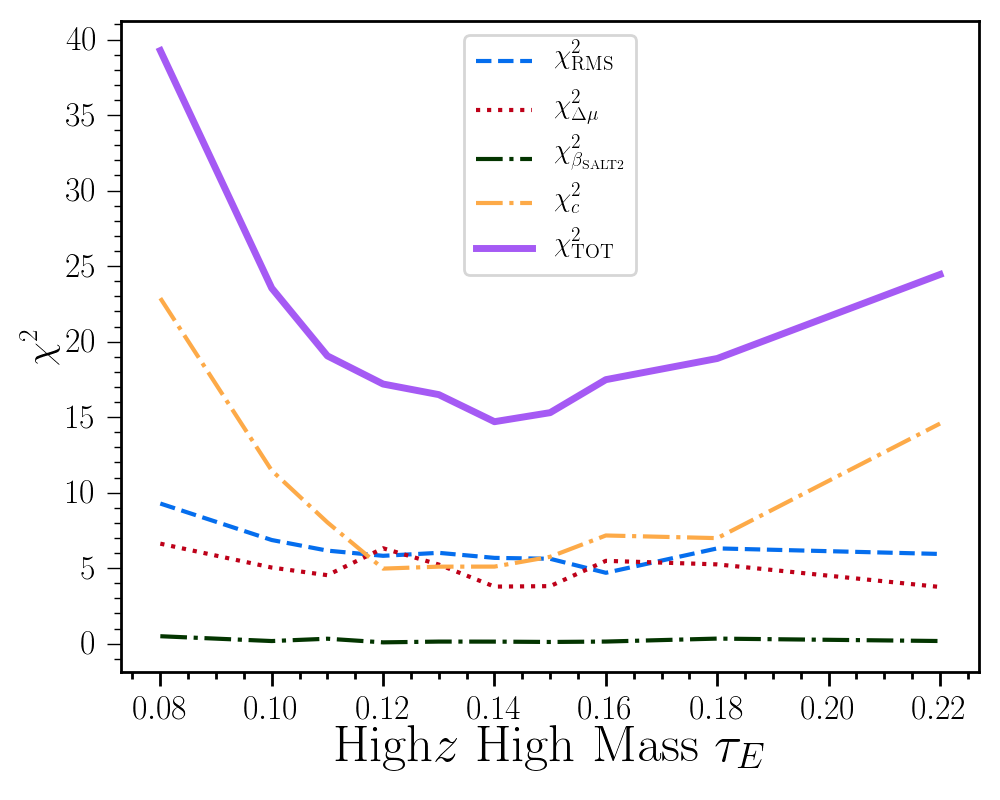} &
 \includegraphics[width=0.23\textwidth]{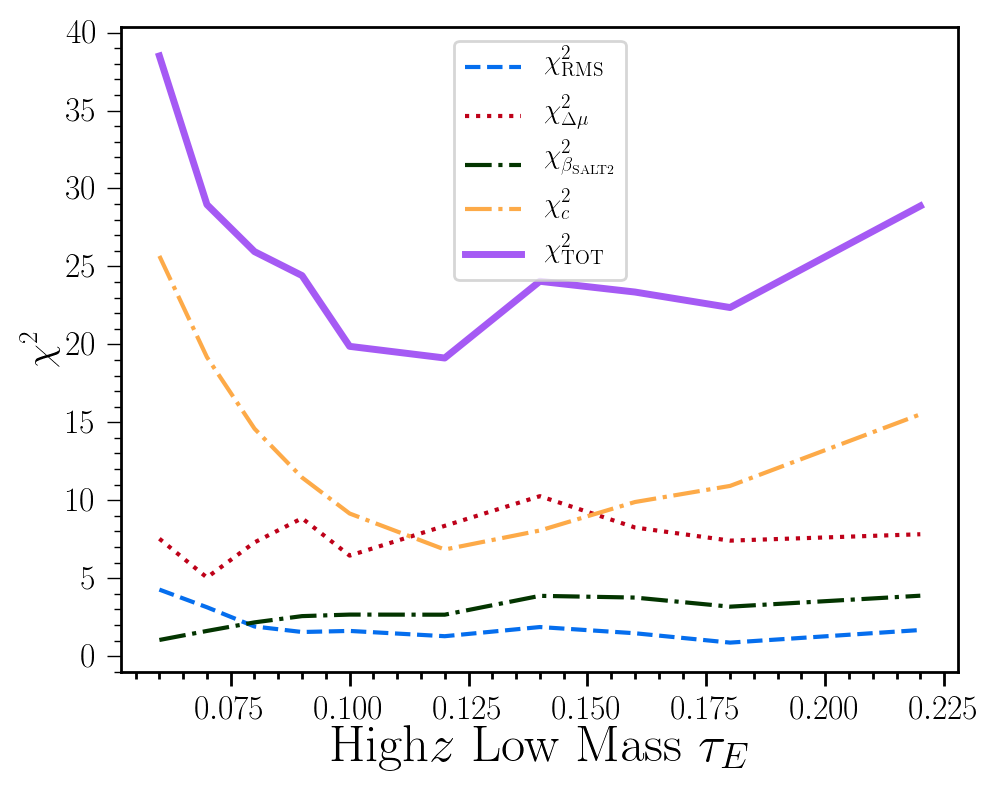} &
    \includegraphics[width=0.23\textwidth]{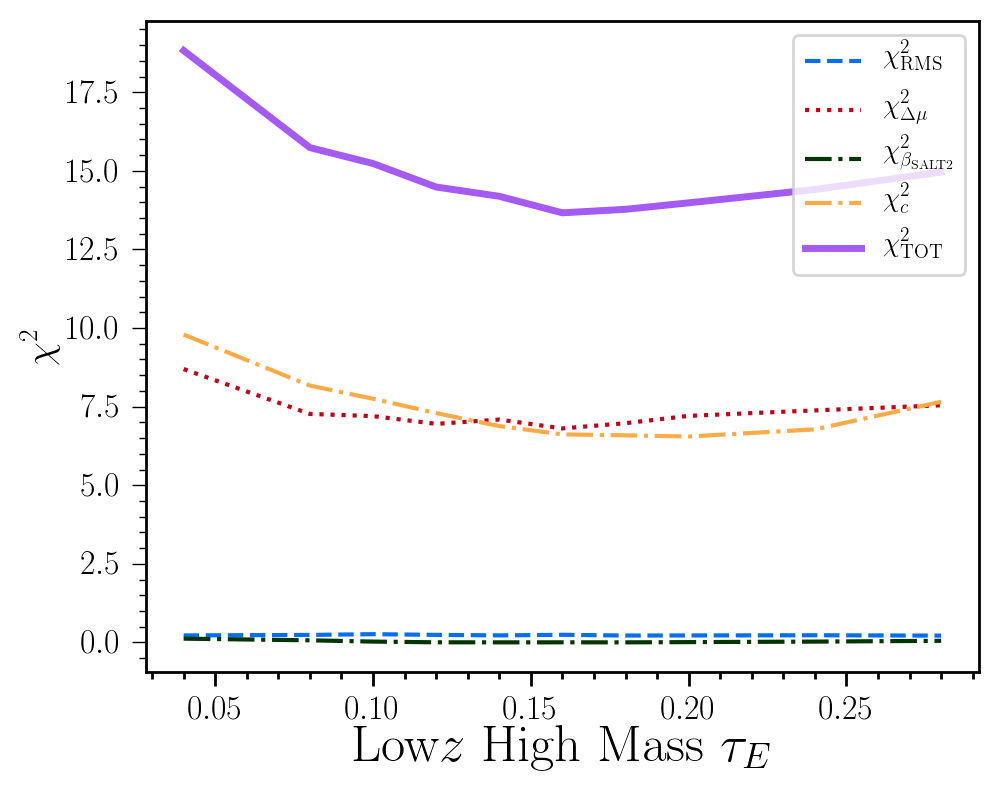} &       \includegraphics[width=0.23\textwidth]{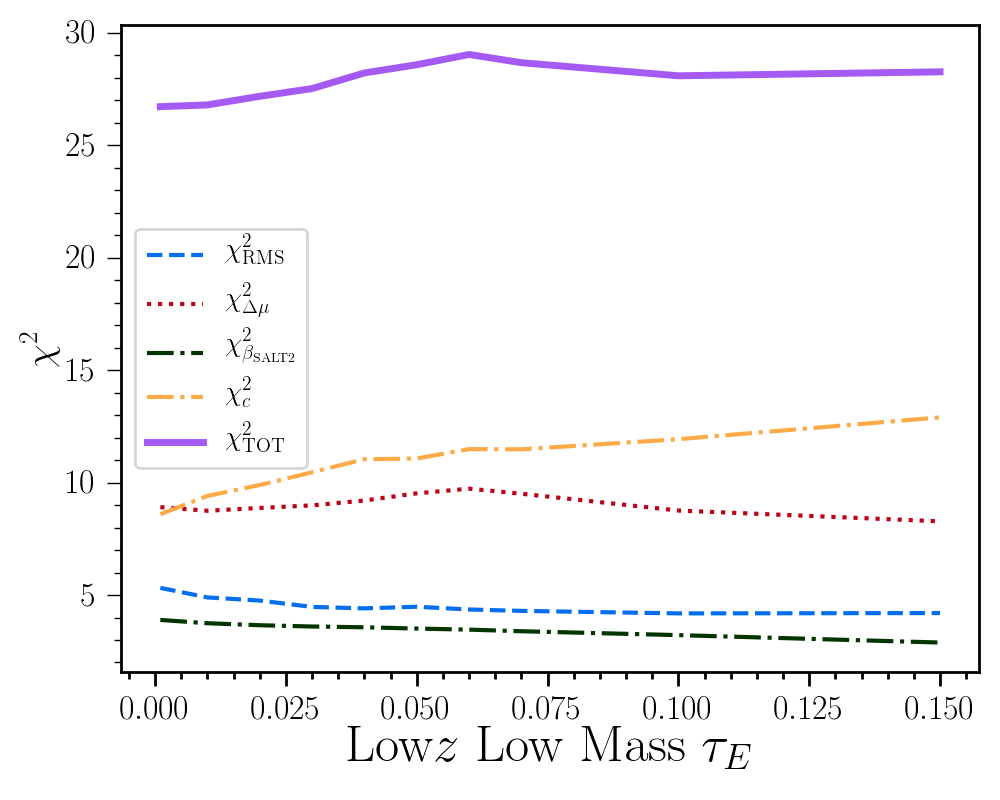} \\
\end{tabular}
\caption{$\chi^2$ surfaces for each of the metrics in Equations \ref{chisqc}, \ref{chisqrms},\ref{chisqdmu}, \ref{chisqbeta}, \& \ref{chisqtot} are shown for each of the fitted parameters in the BS20 model. Each parameter is varied separately with all other parameters held at their best fit values. }
\label{fig:chigrid}
\end{figure*}

\subsection{A1. Model-data Agreement and Parameter Sensitivity}
\label{appendixmeancolor}

Here we review variants on parameters in the various color models to show what impact it has on the three metrics. We list those variants here:
\begin{itemize}
    \item `BS20' - the main model proposed in this work.
    \item `No Dust' - a model with a narrow intrinsic color distribution and a weak ($\bar{\beta}_{SN}=2)$ correlation between color and luminosity.
    \item `Only Dust' - a model with only a dust distribution and a delta function for the intrinsic color distribution.
    \item `G10+SK16' - Described in Section 3.1
    \item `C11+SK16'- Described in Section 3.1
    \item `C11+SK16 + $\bar{\beta}_{SN}$ variation' - a model similar to the `C11+SK16' one, except we allow the $\beta_{SN}$ to vary to reproduce the RMS for redder colors.
    \item `BS20, $\sigma_{\beta_{SN}}=0$' - the nominal BS20 model, except $\beta_{\rm SN}$ values are drawn from a delta function with value $\bar{\beta}_{SN}$. 
    \item `BS20, $R_V+0.5$' - the nominal BS20 model, except we shift our $R_V$ distribution by the full sample by $0.5$.
    \item `BS20, $\tau_E-0.5$' - the nominal BS20 model, except we reduce $\tau_E$ to describe the dust distribution by $0.05$.
    \item `BS20, $\bar{\beta}_{SN}+0.5$' - the nominal BS20 model, except we increase $\bar{\beta}_{SN}$ by 0.5.
    \item `BS20, $\bar{\beta}_{SN}=0$' - the nominal BS20 model, except we set $\beta_{SN}$ to be 0. This effectively describes the intrinsic color distribution as color scatter, similar to what is in the C11+SK16 model. 
    \item `BS20, No $R_V$ variation' - the nominal BS20 model, except the variation in $R_V$ is removed.
\end{itemize}

\begin{figure}
\centering
\includegraphics[width=.95\textwidth]{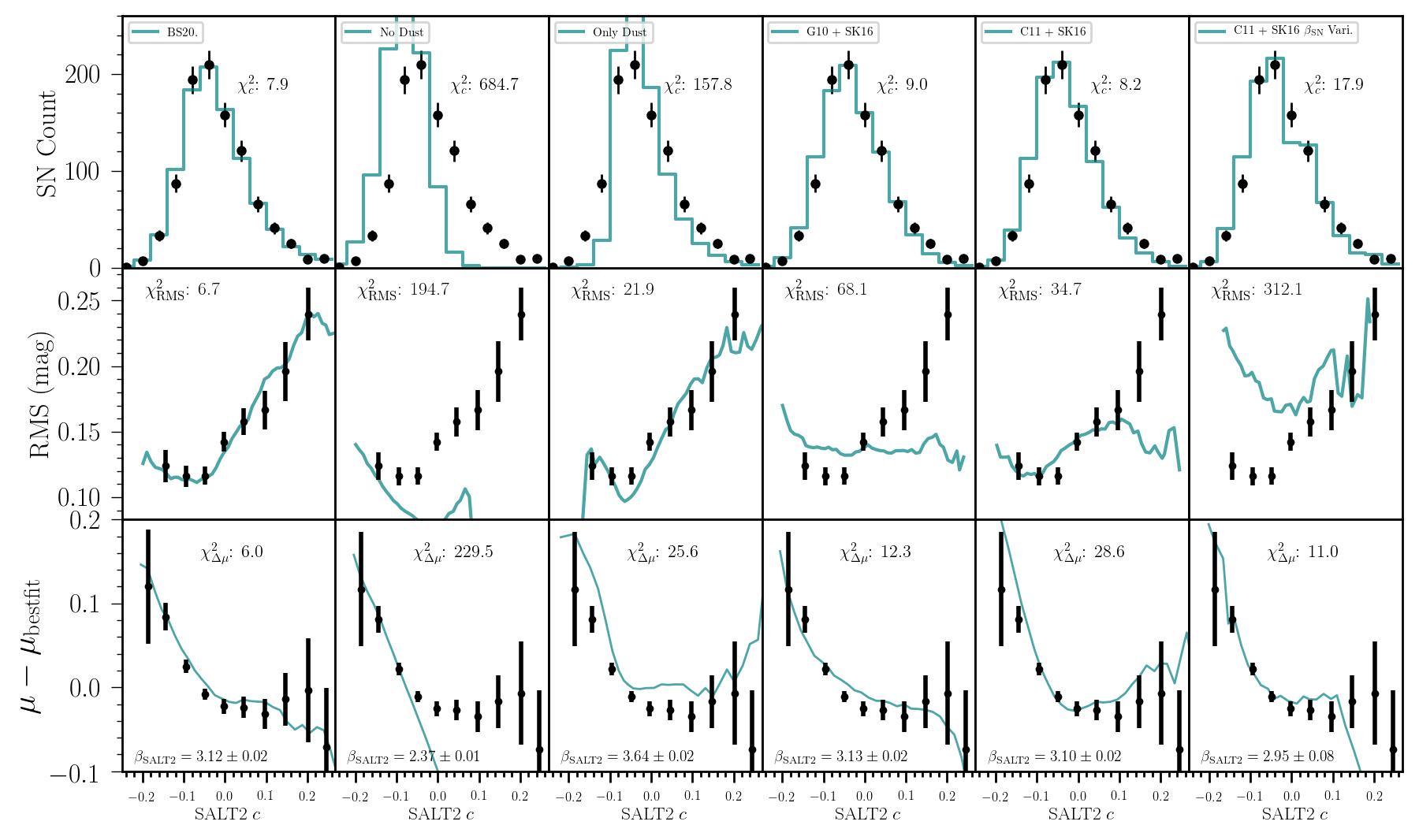}
\includegraphics[width=.95\textwidth]{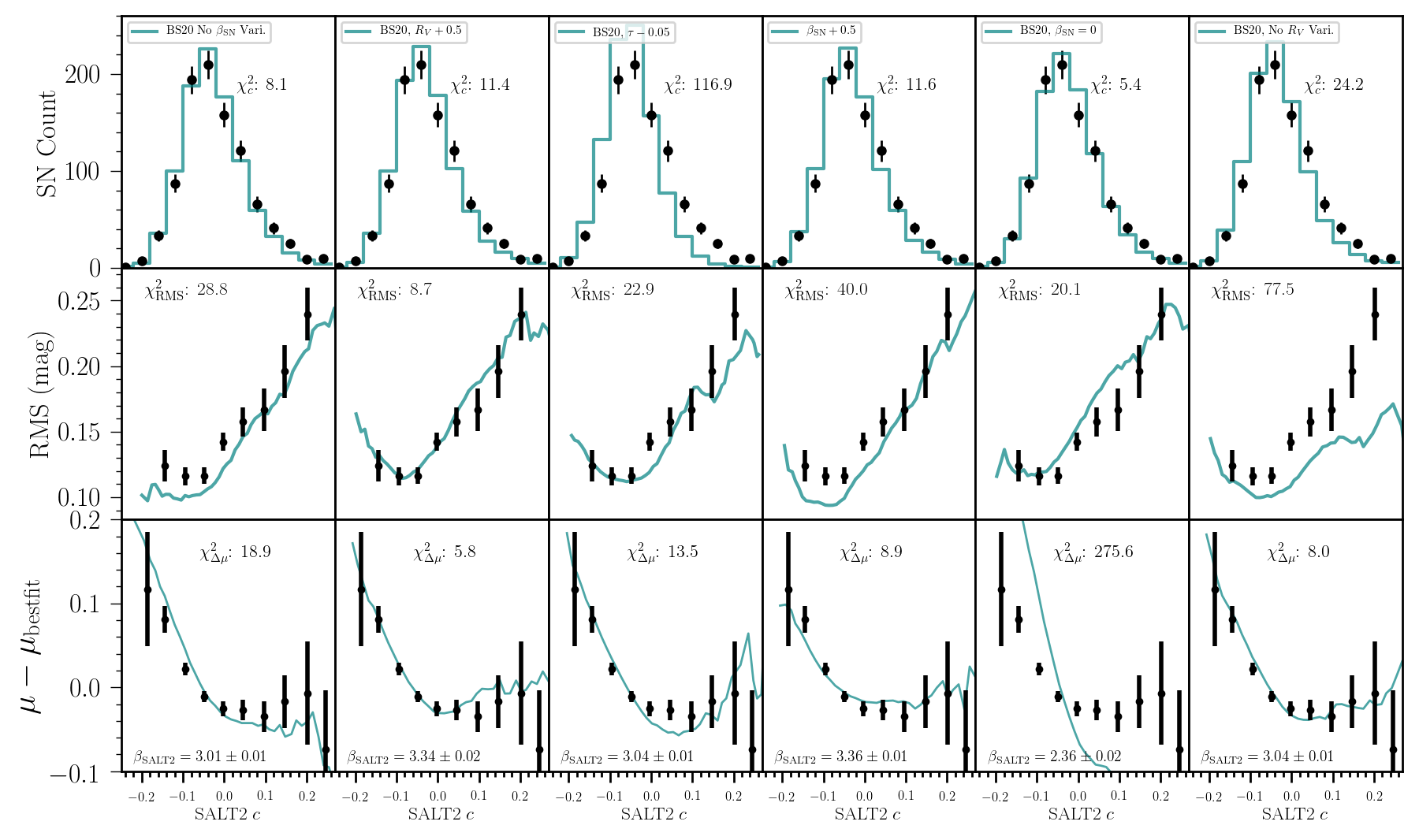}
\caption{Similar to what is shown in Fig.~\ref{fig:chist} and Fig.~\ref{redscatter}, here we show the agreement between data and simulations for variants on our main BS20 Mass-split model as well as for the G10+SK16 and C11+SK16 based models. The $\chi^2$ for each metric is given, and in the bottom panel, we show the recovered $\beta_{SALT2}$ to be compared with that from the data ($\sim3.05$).  The sensitivity to these variations shows the high constraining power of the metrics.}
\label{grid}

\vspace{.1in}

\end{figure}

We show the results from using these different variants in Fig.~\ref{grid}. We include on the bottom panel the recovered $\beta_{SALT2}$ for each case because as some variants may have a good $\chi^2$ in the three metrics, the recovered $\beta_{SALT2}$ is far from that the data ($\sim3.05$).
It is important to note that besides the BS20, G10+SK16 and C11+SK16 models, none of the other models are fit to match the data.

\subsection{A2. Observed Correlations with SALT2 $x_1$}
\label{sec:x1modeling}

\begin{figure*}
\begin{tabular}{ccc}
  \includegraphics[width=0.32\textwidth]{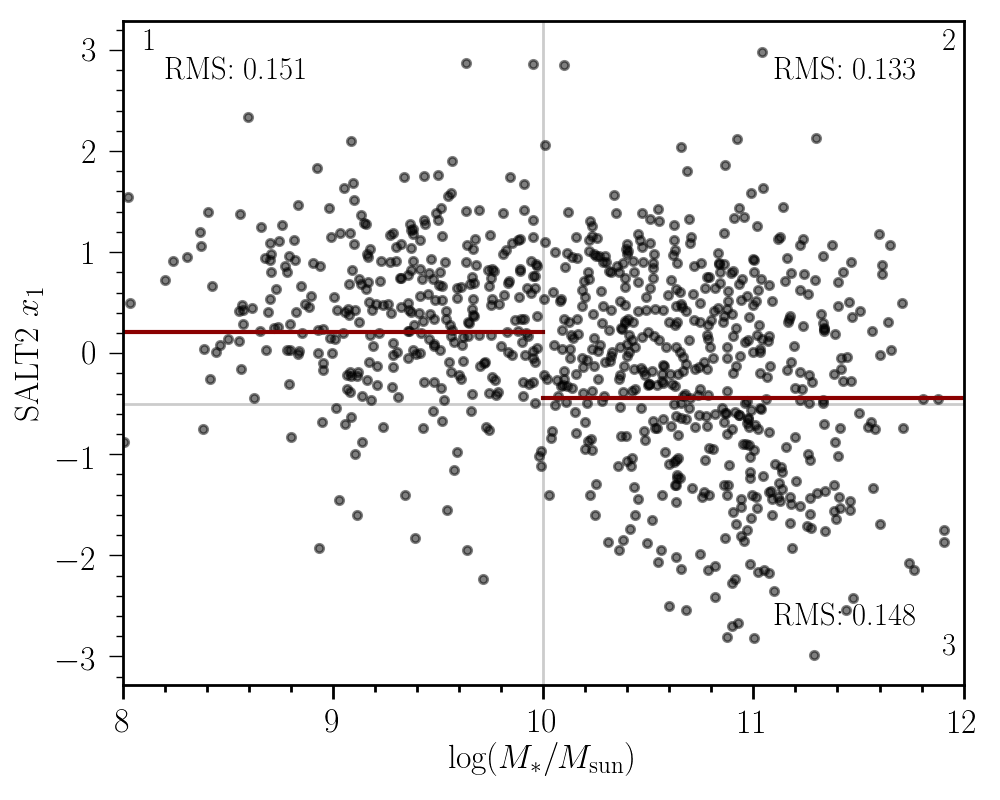} &   \includegraphics[width=0.32\textwidth]{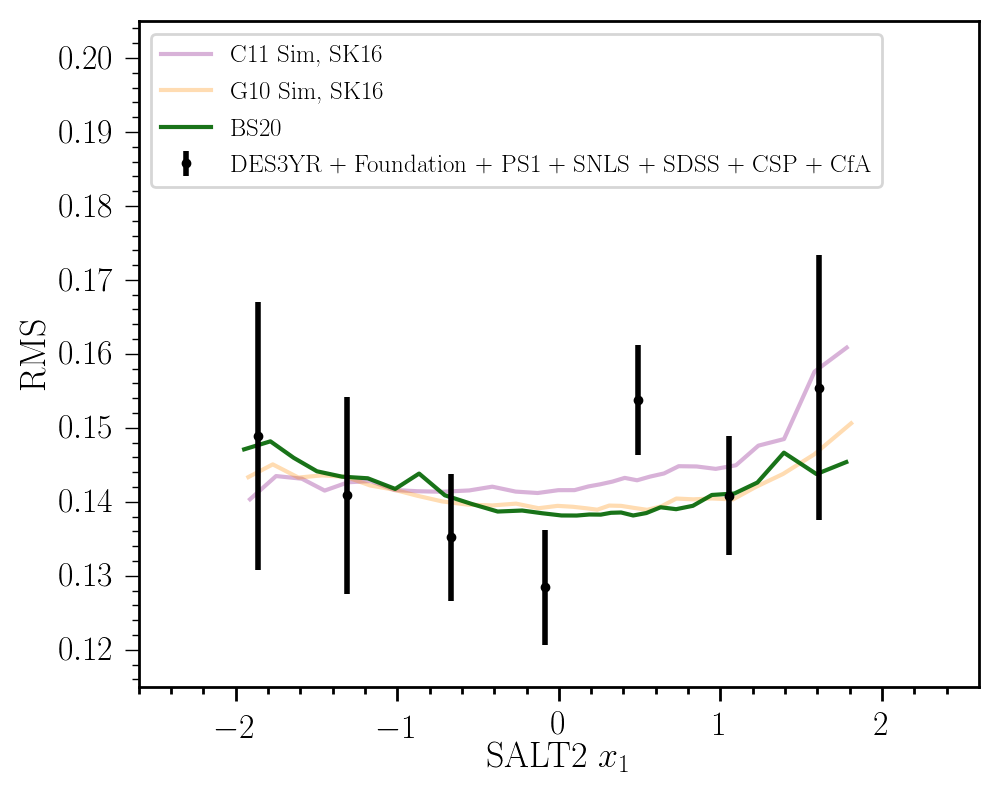} &
  \includegraphics[width=0.32\textwidth]{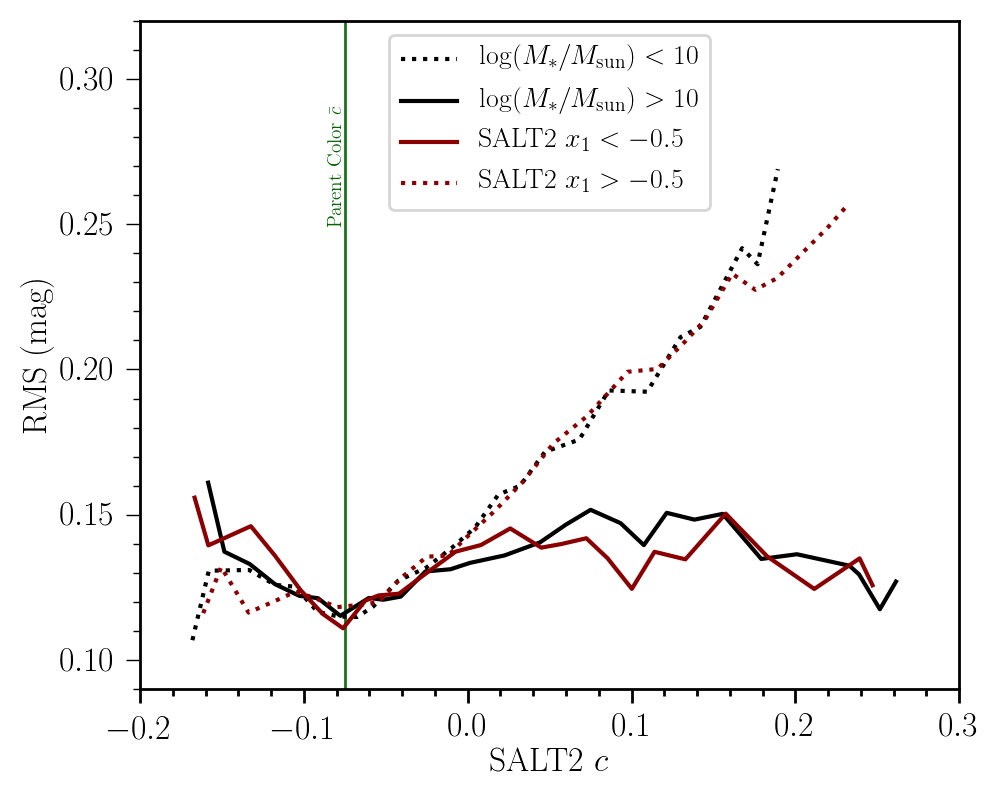} \\
\phantom{aaaa}(a)  & \phantom{aaaaaaa}(b) & \phantom{aaaaaaa}(c)  \\[6pt]
  \includegraphics[width=0.32\textwidth]{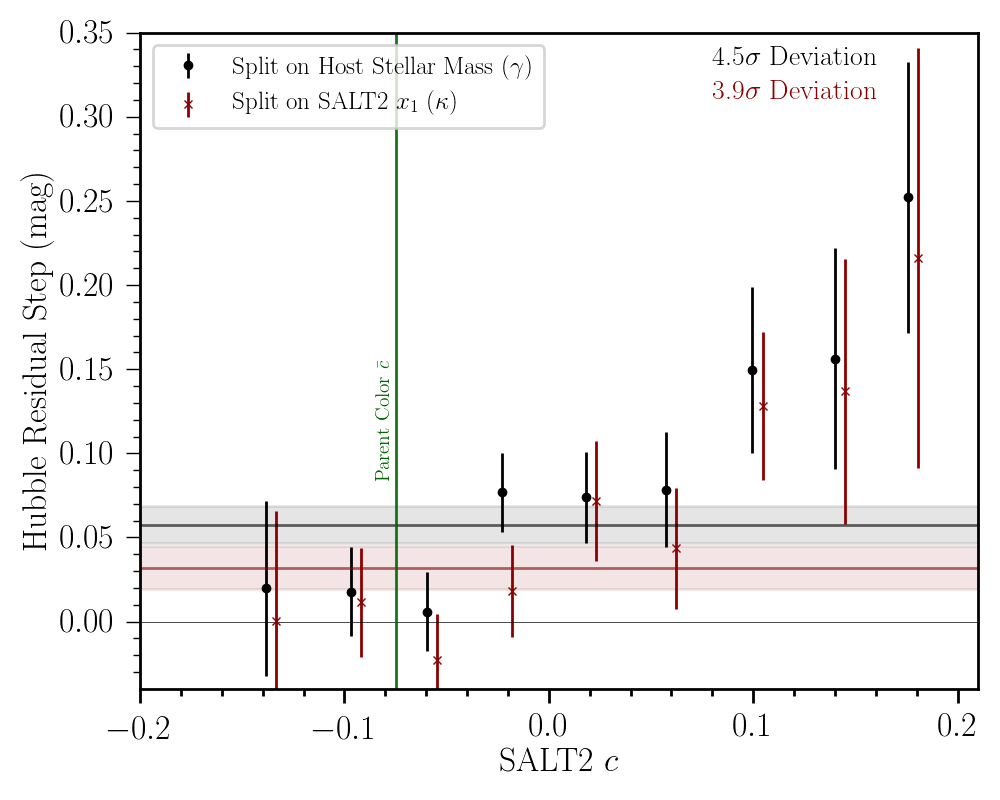} &   \includegraphics[width=0.32\textwidth]{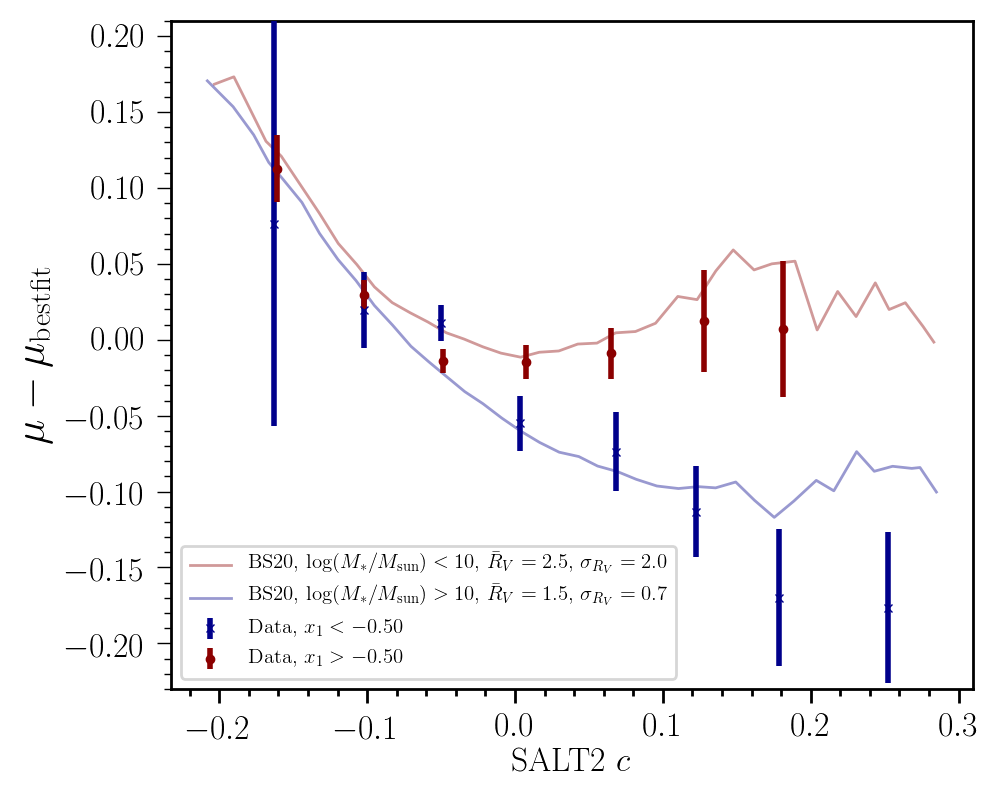} &
  \includegraphics[width=0.32\textwidth]{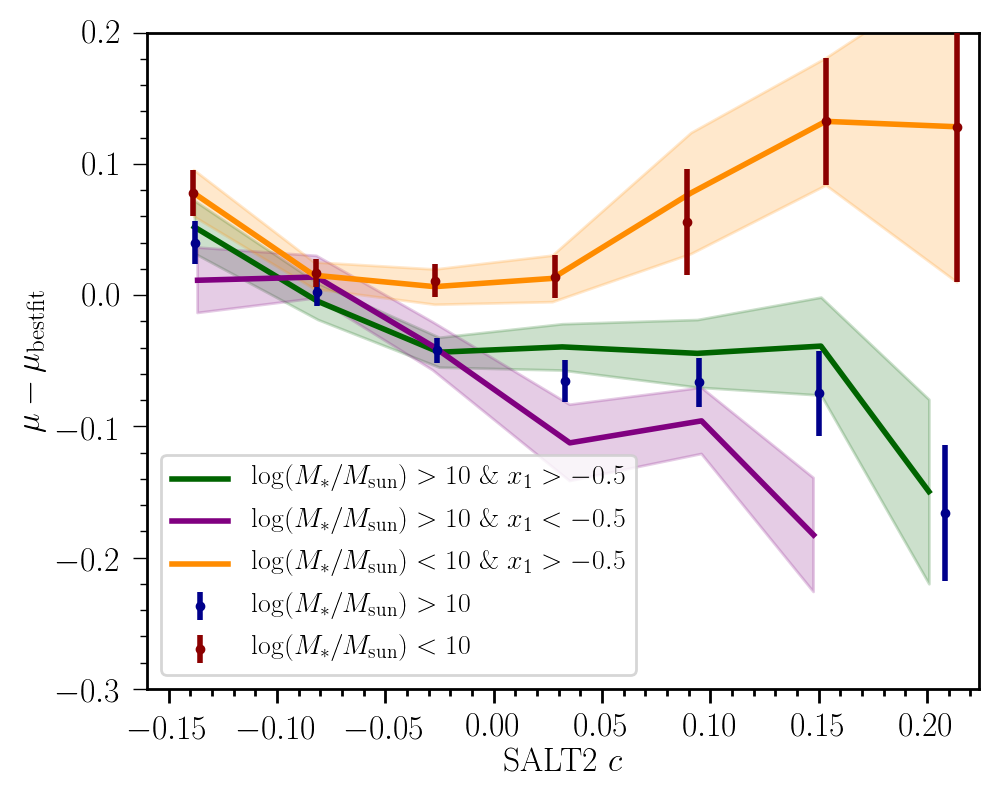} \\
  \phantom{aaaa}(d)  & \phantom{aaaaaaa}(e) & \phantom{aaaaaaa}(f)  \\[6pt]

\end{tabular} 
\caption{\textbf{a)} The correlation between observed SALT2 $x_1$ and host-galaxy stellar mass.  The trend shows a difference in weighted average $x_1$ values (red) when split on $\mathcal{M}_{\rm step}=10$. Hubble residual zero-mean RMS values are reported for subsets of the data. \textbf{b)} RMS of Hubble diagram zero-mean residuals versus SALT2 $x_1$. No dependence is seen. \textbf{c)} RMS of Hubble diagram residuals versus SALT2 $c$ when splitting on both log$(M_*/M_{\rm sun})$ (black) and SALT2 $x_1$ (red). \textbf{d)} Host Mass step as a function of observed color now with $\gamma$ (black) and $\kappa$ (red) overlaid. Simple averages and 1$\sigma$ uncertainties are shown with horizontal lines. Significance of deviation from the respective horizontal lines is reported in text. \textbf{e)} Binned Hubble diagram residuals for the dataset when splitting on $x_1=-0.5$ (points). The predictions using dust and $R_V$ distributions from Sec. \ref{sec:hostmodeling} are overlaid (lines). \textbf{f)} Binned Hubble diagram residuals from three sectors of the dataset (lines) corresponding to the three quadrants shown in Panel (a). Overlaid are the binned Hubble diagram residuals used when only splitting on mass. }
\label{x1fig}
\vspace{.3in}
\end{figure*}

 SN Ia cosmology analyses that measure correlations between SN light curve parameters and host galaxy mass regularly find a correlation between host-galaxy stellar mass and $x_1$ (e.g., \citealp{Sullivan2010,scolnicsys}). This correlation is shown in Fig.~\ref{x1fig}a for our compiled dataset and from this, we expect similar trends with $x_1$ that we observed with host stellar mass in Section 4.  While there is no dependence of the RMS of distance modulus residuals on $x_1$ (Fig. \ref{x1fig}b) seen in the data or predicted from simulations, we do see similar trends with color when splitting on $x_1$ (Fig. \ref{x1fig}c) as we do when splitting on $\mathcal{M}_{\rm host}$. We also compute a Hubble residual step when splitting on $x_1$:
\begin{equation}
\label{Eq:x1step}
\delta\kappa = \kappa \times [1+e^{(x_1-x_{1_{\rm step}})/0.01}]^{-1}-\frac{\kappa}{2},
\end{equation}
at SN~Ia stretch step location ($x_{1_{\rm step}}$) of -0.5 is assumed, albeit $x_{1_{\rm step}}$ values between -0.5 and +0.5 provide good discrimination between sub-samples according to our three metrics. We determine $\kappa$ for the sample in discrete color bins (Fig.~\ref{x1fig}d). When deriving $\delta \kappa$ for the full sample with a single $x_{1_{\rm step}}$ split, we find $\delta \kappa = 0.032\pm.011$ mag, roughly half the size of the step when splitting by host stellar mass. As shown in Fig.~\ref{x1fig}d, similarly to host mass, the magnitude of $\kappa$ depends on color; there is a $3.9\sigma$ deviation relative to a single step. 

When examining Hubble diagram residual biases in bins of color (Fig.~\ref{x1fig}e), simulations using the dust and $R_V$ distributions that were fit in Sec. \ref{sec:hostmodeling} roughly predict the residuals when splitting on $x_1$. This indicates that  $x_1$ and $\mathcal{M_{\rm host}}$ yield similar information about the dust properties. However, upon studying the mean Hubble residual bias with color, as shown in Fig.~\ref{x1fig}f, we find the information from $x_1$ and $\mathcal{M_{\rm host}}$ are complementary in potentially constraining $R_V$ as the difference in Hubble residuals from the subsample of low $x_1$ values and large host mass values (purple) in comparison to those from high $x_1$ values and small mass values (orange) is larger than simply splitting on host mass (data points) as was done in Section 4. This finding is consistent with studies like \cite{Rose_2019}, which argue that combinations of various host-galaxy properties and light-curve parameters could further improve the standardizability of SNe Ia brightnesses, as well as with \cite{Galbany12} who find that $x_1$ is a good discriminator of galaxy morphology.

\subsection{A3. The SALT2 Color Law}

\begin{figure}
\centering
\includegraphics[width=.44\textwidth]{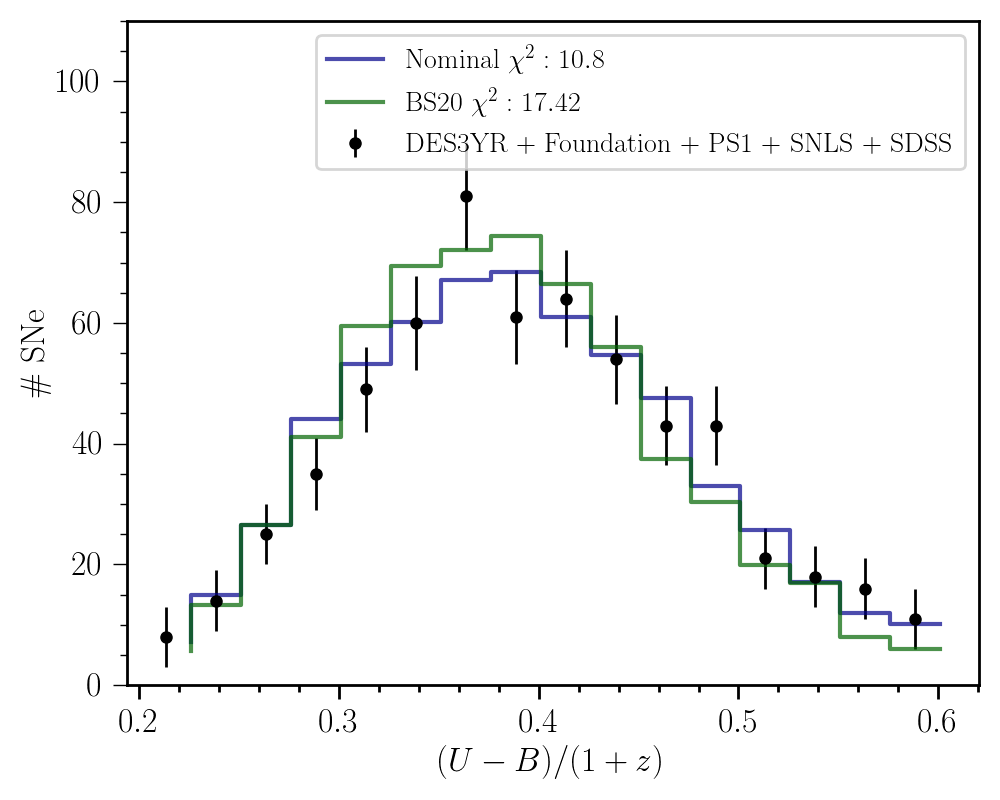}
\includegraphics[width=.44\textwidth]{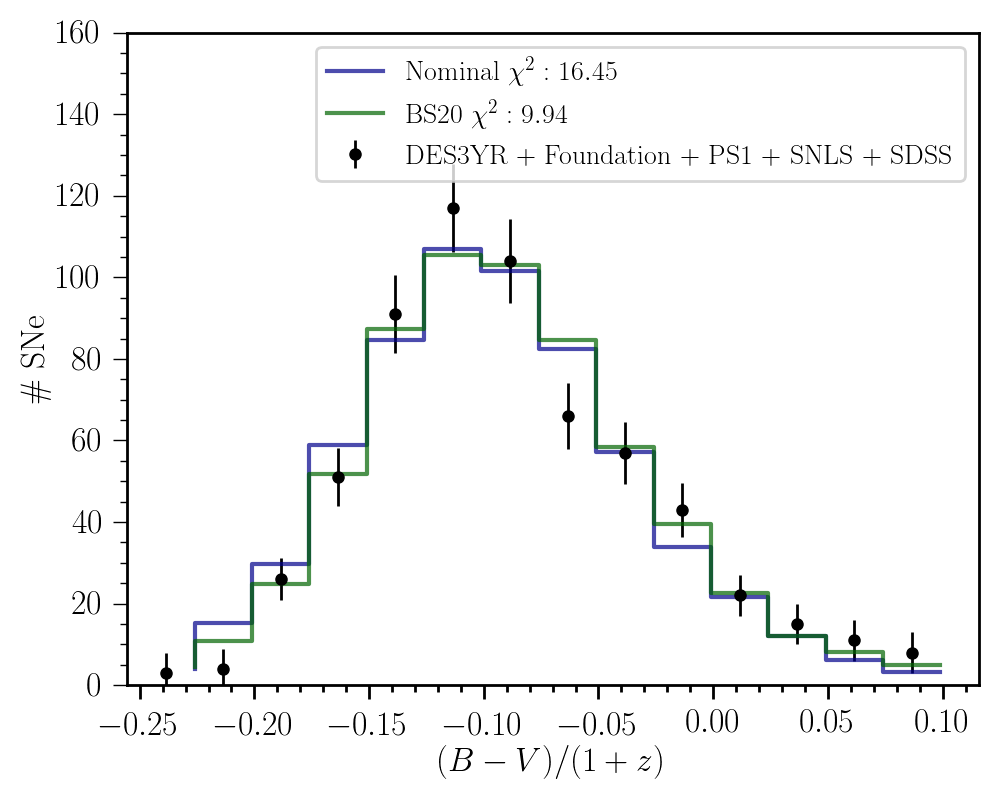}
\caption{Predicted distribution of restframe colors in $(U-B)/(1+z)$ and $(B-V)/(1+z)$ from BS20 and the G10+SK16 model (Nominal), where $z$ is the observed redshift of the SNe. The $\chi^2$ defined in Eq. 3 is reported. Lesser agreement in $(U-B)$ for the BS20 model motivates retraining of the light-curve model in a future study.}
\label{fig:restframe}
\vspace{.3in}
\end{figure}

In the discussion, we explain that a future analysis should retrain the color law(s) to match the data, rather than rely on our a-posteriori model selection. In Fig. \ref{fig:restframe}, we derive the predicted distribution of peak rest-frame colors from our model, and from a nominal SALT2 based color model (G10+SK16) and compare to data. This is done by k-correcting observations to the rest-frame using the SALT2 spectral model. We show that the BS20 model better predicts the observed $(B-V)$ distribution while G10+SK16 based better predicts the observed $(U-B)$ distribution, both by similar amounts in $\chi^2$. As the BS20 model selection had little sensitivity to the rest-frame UV colors, this evaluation is not surprising. It is moderately surprising, however, that given the lack of UV sensitivity in the metrics, the BS20 does as well as it does in the UV. Still, we argue that a proper retraining of the light-curve model that incorporates flexibility for discrimination between the intrinsic SN~Ia color law and dust color laws is needed in the future. Interestingly, \cite{Amanullah15} shows that in the UV, a Fitzpatrick $R_V=2.2$ matches observations of nearby SNe significantly better than the SALT2 color law. As such, we expect that retraining based on our model can improve the plot shown here.

\newpage 

\subsection{A4. Similar but non-viable model.}

\begin{figure}
\centering
\includegraphics[width=0.49\textwidth]{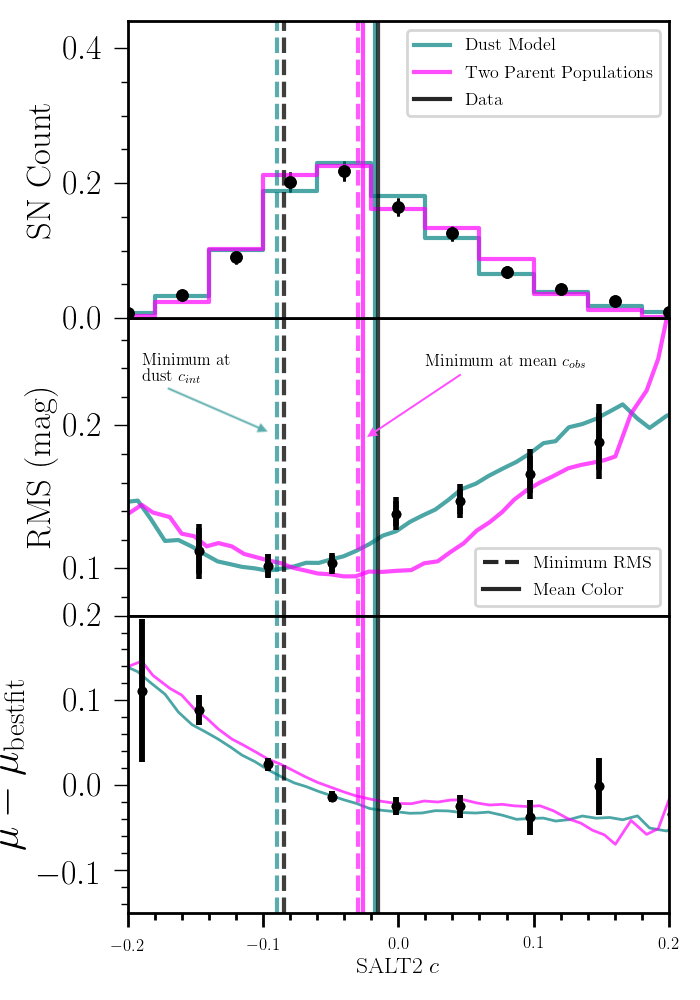}
\caption{Same as Fig. \ref{fig:cartoon2} but with the inclusion of a multiple-color population model (pink). The multiple-color population model reproduces the trends observed in the data for the color distribution (top) and Hubble diagram residuals (bottom), but does not reproduce the scatter in the Hubble diagram as a function of color (middle). The minimum scatter in the data (vertical black dashed line) and in the dust model (vertical teal dashed line) occurs bluer than the mean observe colors (black/teal solid vertical lines). However this is not the case for the multiple color population model, which has minimum scatter (pink dashed vertical line) occurring at the mean observed color (pink solid vertical line).}
\label{fig:twopops}
\end{figure}

 Recent studies proposed that are more than one parent populations of SNIa color  \cite{Milne2013,Pan_2014,Stritzinger2018,Jiang2018,Bulla2020,Kelsey2020,Gaitan2020} with differing $\beta$'s. To illustrate, we follow the optimization process described in Section \ref{sec:fittingmethod} and fit for parameters of a two component color population model. We assume that SNIa colors are drawn from a bluer Gaussian ($c^{\rm blue}_{\rm int}$, $\sigma_c^{\rm blue}$) and a redder Gaussian ($c^{\rm red}_{\rm int}$, $\sigma_c^{\rm red}$). A larger $\sigma_c^{\rm red}$ in comparison to $\sigma_c^{\rm blue}$ results in the asymmetric histogram seen in the data (Fig. \ref{fig:twopops} top). Each color population is given its own color-luminosity relation ($\beta^{\rm blue}$ \& $\beta^{\rm red}$) to explain the color dependent residuals in the Hubble diagram (Fig. \ref{fig:twopops} bottom). In order to explain the scatter floor in the data (Fig. \ref{fig:twopops} middle), we must include a $\sigma_{\rm coh}$ term and to explain the increasing RMS as a function of color we allow for $\sigma_\beta^{\rm blue}$ \& $\sigma_\beta^{\rm red}$. 

A simulation with the best-fit population parameters is shown in the pink curves of Figure \ref{fig:twopops}. We find that while the color distribution (top) and the trend of Hubble residuals with color (bottom) are reproduced, we are not able reproduce the trend of RMS scatter with color (middle). We note that for the multiple population model, the color at which Hubble diagram scatter is smallest occurs at the mean observed color. This is the expectation arising from the assumption of two symmetric parent populations. However this is not what is observed in the data. In the data, the color corresponding to the minimum scatter in Hubble diagram is -0.085, far bluer than the mean observed color (-0.016); this is indicative of one-sided color variance (i.e. dust). Our best-fit dust model is shown for comparison (teal) and agrees with this expectation.

Furthermore, we note that this multiple-population model is more complicated in its physical underpinnings compared to a model where there is an intrinsic color and dust.  For this new model, there remains no unanimously accepted reason for why there would be two populations of SNe with differing colors and differing color-luminosity relations. Additionally, this model has one more parameter (9) than the model we propose (8), and despite its added complexity the agreement with the observed data is poorer. 

Lastly, as is explained in Section \ref{sec:hostmodeling}, an additional correlation between dust-model parameters ($R_V$) and host properties can fully explain the mass step. Likewise, if one is motivated to use the multiple population model to explain the mass step, additional complexity must be added. From Fig. 6b, there are clearly 3 separate slopes in the Hubble residuals versus color when split on host-galaxy mass.  Therefore, in the multiple population methodology, this implies that 3 separate populations of SNe with three $\beta$'s would be needed to explain the mass step. Bi-modal populations motivated by  two different progenitor scenarios are inconsistent with this finding of three separate slopes.

\bibliographystyle{apj}
\bibliography{sample}

\renewcommand{\thefigure}{A.\arabic{figure}}

\setcounter{figure}{0}

\end{document}

%% file: params.tex
\begin{table*}
\label{ptable}
\noindent
\caption{Parameters used for BS20 model.}

\makebox[.99\textwidth]{
\resizebox{.89\textwidth}{!}{
\begin{tabular}{|c|c|c|c|c|c|c|c|c|}
\hline\hline
Model & Sample &$\bar{c}$  & $\sigma_c$ & $\bar{\beta}_{\rm SN}-1$ & $\sigma_{\beta_{\rm SN}}$  & $\bar{R}_V$ & $\sigma_{R_V}$ & $\tau_E$  \\
\hline

No-Mass-split:&&&&&&&&\\
Full& CfA, CSP, Foundation &
{-0.084}$\pm$0.004&
{0.042}$\pm$0.002&
{0.98}$\pm$0.18&
{0.35}$\pm$0.20&
2.0$\pm$ 0.2&
1.4$\pm$ 0.2&
0.17$\pm$ 0.04
\\

Full & DES, PS1, SNLS, SDSS &
{-0.084}$\pm$0.004&
{0.042}$\pm$0.002&
{0.98}$\pm$0.18&
{0.35}$\pm$0.20&
2.0$\pm$ 0.2&
1.4$\pm$ 0.2&
0.10$\pm$ 0.02
\\
\hline
Mass-split:&&&&&&&&\\

\phantom{.}High-mass\footnote{High mass: Host log$(M_*/M_{\rm sun}) > 10$} & CfA, CSP, Foundation&
{-0.084}$\pm$0.004&
{0.042}$\pm$0.002&
{0.98}$\pm$0.18&
{0.35}$\pm$0.20&
{1.50}$\pm$0.25&
{1.3}$\pm${0.2}&
{0.19}$\pm${0.08}
\\

High-mass & DES, PS1, SNLS, SDSS   &
{-0.084}$\pm$0.004&
{0.042}$\pm$0.002&
{0.98}$\pm$0.18&
{0.35}$\pm$0.20&
{1.50}$\pm$0.25&
{1.3}$\pm${0.2}&
{0.15}$\pm${0.02}
\\

\phantom{.}Low-mass\footnote{Low mass: Host log$(M_*/M_{\rm sun}) < 10$} & CfA, CSP, Foundation &
{-0.084}$\pm$0.004&
{0.042}$\pm$0.002&
{0.98}$\pm$0.18&
{0.35}$\pm$0.20&
{2.75}$\pm$0.35&
{1.3}$\pm${0.2}&
${0.01}^{+0.05}_{-0.01}$
\\

Low-mass & DES, PS1, SNLS, SDSS   &
{-0.084}$\pm$0.004&
{0.042}$\pm$0.002&
{0.98}$\pm$0.18&
{0.35}$\pm$0.20&
{2.75}$\pm$0.35&
{1.3}$\pm${0.2}&
{0.12}$\pm${0.02}
\\
\hline

\end{tabular}%
}}

\vspace{.03in}

\label{tab:paramtable}
\end{table*}

%% file: chisq.tex
\begin{table}
\noindent
\caption{$\chi^2$ for each evaluated model.}

\makebox[.49\textwidth]{%
\resizebox{.48\textwidth}{!}{
\begin{tabular}{|c|c|c|c|c|c|c|c|}
\hline\hline
Scatter Model & Color Model & $\chi^2_c$ & $\chi^2_{\rm RMS}$ & $\chi^2_{\Delta \mu}$& $\chi^2_{\beta_{\rm SALT2}}$ & $\chi^2_{\rm Tot}$ & \# of Parameters\footnote{Note: \# of parameters is counted in the text}
 \\
\hline
G10 & SK16 & 22.0 & 68.1 & 12.3 & 1.6 & 104.0 & 9  \\
C11 & SK16 & 19.4 & 34.7 & 28.6 & 1.3 & 84.0 & 9\\
BS20 No-Mass-split & BS20 &  7.9 & 6.7 & 6.0 & 1.7 & 22.3 & 8  \\
\hline

\end{tabular}%
}}

\vspace{.02in}

\label{tab:chisq}
\end{table}

%% file: wtable.tex
\begin{table*}
\label{wtable}
\noindent

\caption{Results from Large $\Lambda$CDM Simulations and 1445 SNe~Ia}

\centering

\begin{tabular}{|c|c|c|c|c|c|}
\hline\hline
SN~Ia Color Model & Host Dust Model & SN~Ia Color Model & Host Dust Model  & SN~Ia + Dust  & $w$CDM + Planck '16 \\
 Data\footnote{ Datasets are based on large simulations of $\sim$10,000 SNe~Ia. Each dataset (row) is a unique statistical realization.} & Data & 1D BiasCor\footnote{ Bias Correction samples are large simulations of $>$1,000,000 SNe~Ia.} & 1D BiasCor  & $\beta_{\rm SALT}$ & $\Delta w$\footnote{$\Delta w$=$w_{\rm fit}-w_{\rm BS20}$: this is relative to the last row (BS20) of each dataset grouping.} \\
\hline
BS20 & BS20 & C11 + SK16 Parent & No Host Dust & $3.07\pm0.01$ &   -0.04\\
BS20 & BS20 & G10 + SK16 Parent & No Host Dust & $3.09\pm0.01$ &   -0.03\\
BS20 & BS20 & BS20 & BS20 & $3.12\pm0.01$ &  0.00 \\ 
&&&&& \\
Real Data & Real Data & C11 + SK16 Parent & No Host Dust  & $3.06\pm0.06$ & -0.04  \\
Real Data & Real Data & G10 + SK16 Parent & No Host Dust  & $3.05\pm0.06$ & -0.03   \\
Real Data & Real Data & BS20 & BS20  & $3.06\pm0.06$ &   0.00  \\
\hline

\end{tabular}%

 \vspace{.02in}

\end{table*}